\RequirePackage{amsmath}
\documentclass[runningheads]{llncs}
\usepackage{qtree}
\usepackage{tree-dvips}
\usepackage{amsmath}
\usepackage[T1]{fontenc}
\usepackage{graphicx}
\usepackage{tikz}
\usepackage{amsfonts}
\usepackage{float}
\usepackage{xcolor}
\usepackage{mathabx}
\usepackage{hyperref}
\usepackage{multirow}

\usepackage[noend,ruled,linesnumbered]{algorithm2e}
\newcommand{\ovrt}{\ominus}
\newcommand{\ohrz}{\mathbin{\rotatebox[origin=c]{90}{$\ovrt$}}}
\newcommand{\no}[1]{}
\newcommand{\vir}[1]{``#1''}
\newcommand{\zero}{\texttt{0}}

\newcommand{\one}{\texttt{1}}

\newcommand{\syma}{\texttt{a}}
\newcommand{\symb}{\texttt{b}}

\newcommand{\dd}{\mathinner{.\,.}}
\newcommand{\Mmn}{\mathcal{M}_{m\times n}}
\newcommand{\Mnn}{\mathcal{M}_{n\times n}}
\newcommand{\MathM}{\mathcal{M}}
\newcommand{\mxn}{[1\dd m] \times [1\dd n]}
\newcommand{\gexp}{\mathtt{exp}}
\newcommand{\rowlin}{\mathtt{rlin}}
\newcommand{\phlin}{\mathtt{phlin}}

\newcommand{\map}{\mathtt{map}}
\newcommand{\LS}{\mathtt{ls}}
\newcommand{\RS}{\mathtt{rs}}
\newcommand{\US}{\mathtt{us}}
\newcommand{\DS}{\mathtt{ds}}

\newcommand{\deltaCM}{\delta_\square}
\newcommand{\gammaCM}{\gamma_\square}
\newcommand{\bCM}{b_\square}
\newcommand{\nn}{\mathbf{n}}
\newcommand{\MathMM}{\MathM_\nn}
\newcommand{\Dim}{d}
\newcommand{\Left}{\mathtt{L}}
\newcommand{\Right}{\mathtt{R}}
\newcommand{\Up}{\mathtt{U}}
\newcommand{\Down}{\mathtt{D}}

\usetikzlibrary{positioning, matrix, arrows}
\usepackage{ifthen}
\usepackage{comment}

\newcommand{\ignore}[1]{}

\definecolor{darkpastelgreen}{rgb}{0.01, 0.5, 0.24}
\long\def\fullversion#1{{#1}}

\begin{document}

\title{Generalization of Repetitiveness Measures for Two-Dimensional Strings}

\author{Lorenzo Carfagna\inst{1}\orcidID{0009-0005-9591-057X} \and\\
Giovanni Manzini\inst{1}\orcidID{0000-0002-5047-0196}\and\\
Giuseppe Romana\inst{2}\orcidID{0000-0002-3489-0684}\thanks{Corresponding author.}
\and\\
Marinella Sciortino\inst{2} \orcidID{0000-0001-6928-0168}\and\\
Cristian Urbina\inst{3,4}\orcidID{0000-0001-8979-9055}}

\institute{Dipartimento di Informatica, University of Pisa, Italy \\
\email{lorenzo.carfagna@phd.unipi.it, giovanni.manzini@unipi.it} \and
Dipartimento di Matematica e Informatica, University of Palermo, Italy \\
\email{\{giuseppe.romana01, marinella.sciortino\}@unipa.it}\and
Department of Computer Science, University of Chile, Chile \and Centre for Biotechnology and Bioengineering (CeBiB), Chile \\\email{crurbina@dcc.uchile.cl}}

\authorrunning{L. Carfagna, G. Manzini, G. Romana, M. Sciortino, and C. Urbina}

\maketitle
\begin{abstract}
The problem of detecting and measuring the repetitiveness of one-dimensional strings has been extensively studied in data compression and text indexing. Our understanding of these issues has been significantly improved by the introduction of the notion of {\em string attractor} [Kempa and Prezza, STOC 2018] and by the results showing the relationship between attractors and other measures of compressibility. 


When the input data are structured in a non-linear way, as in two-dimensional strings, inherent redundancy often offers an even richer source for compression. However, systematic studies on repetitiveness measures for two-dimensional strings are still scarce. In this paper we extend to two or more dimensions the main measures of complexity introduced for one-dimensional strings. We distinguish between the measures $\delta$ and $\gamma$, defined in terms of the substrings of the input, and the measures $g$, $g_{rl}$, and $b$, which are based on copy-paste mechanisms. We study the properties and mutual relationships between these two classes and we show that the two classes become incomparable for $d$-dimensional inputs as soon as $d\geq 2$.
Moreover, we show that our grammar-based representation of a $d$-dimensional string of size $N$ enables direct access to any symbol in $O(\log N)$ time.

We also compare our measures for two-dimensional strings with the 2D Block Tree data structure [Brisaboa et al., Computer J., 2024] and provide some insights for the design of future effective two-dimensional compressors. 
\end{abstract}

\section{Introduction}

In the latest decades, the amount of data generated in the world has become massive but it has been observed that, in many fields, most of this data is highly repetitive. For the study of highly repetitive one-dimensional data, an important role is played by the notions of {\em substring complexity},  {\em string attractor}, {\em bidirectional macro scheme}, and {\em grammar compression}, which lead to the definition of the repetitiveness measures $\delta$, $\gamma$, $b$, $g$, and $g_{rl}$ (see \cite{KP18,NavarroSurvey} for the definitions and their remarkable properties). Such measures provide the theoretical basis for the design and analysis of data compressors and compressed indexing data structures for highly repetitive data~\cite{BCGGKNOPTjcss20,Navacmcs20.2}. 

Two-dimensional data, ranging from images to matrices, often contains inherent redundancy, wherein identical or similar substructures recur throughout the dataset. This great source of redundancy can be exploited for compression. Recently, Brisaboa et al. introduced the conceptually simple {2D Block Trees} data structure  to compress two-dimensional strings supporting the efficient access to the individual symbols~\cite{BGGBNdcc18}. Experiments have shown the practicality of 2D Block Trees for storing raster images and the adjacency matrix of Web graphs~\cite{BrisaboaGGN24}.
On the theoretical side, in~\cite{CarfagnaManzini2023} the authors proposed two generalizations of the measures $\delta$ and $\gamma$ for {square} 2D input strings. Such generalized measures are based on properties of the {\em square} submatrices of the input string. The choice of considering only square submatrices was dictated by purely practical considerations: 2D submatrices can be efficiently handled using the 2D Suffix Tree data structure~\cite{siamcomp/Giancarlo95}, and the 2D Block Tree is based on repeated occurrences of square submatrices in the input string. Indeed, \cite{CarfagnaManzini2023,CarfagnaManzini2024} provide the first theoretical analysis of the space usage of the {2D Block Tree} and an optimal linear time construction algorithm. 


In this paper, 
we generalize the 1D measures mentioned above  ($\delta$, $\gamma$, $b$, $g$ and $g_{rl}$)  to 2D strings, considering submatrices of any rectangular shape and not only square submatrices as in~\cite{CarfagnaManzini2023,CarfagnaManzini2024}. Our main results can be summarized as follows:

\begin{itemize}

\item we show that using rectangular submatrices all the above 1D measures can be naturally generalized to 2D strings, and we compare their properties with those of the square-based 2D measures introduced in~\cite{CarfagnaManzini2023,CarfagnaManzini2024}; 

\item we establish some relationships between the new 2D measures and we prove that some properties which are valid in 1D are no longer valid in 2D; other properties are still valid but the gap between some measures can be asymptotically much larger than in 1D;

\item we show that the measures $\delta$ and $\gamma$, which have a simple definition in terms of the submatrices of the input, are not as expressive as in 1D, while the measures $g$, $g_{rl}$ and $b$, which are based on a copy-paste mechanism, appear to retain their role of capturing the repetitiveness of the input even in 2D;

\fullversion{
\item we show that a 2D grammar representing an $m\times n$ string can be enriched with additional information supporting the random access to individual symbols of the original string in {$O(\log mn)$} time.
}

\item we show that the 2D Block Tree data structure, being based on square submatrices, fails to capture the regularities of some inputs which are instead captured by the measures $g$, $g_{rl}$ and $b$;

\item we use our generalized measures to analyze a frequently used heuristics for 2D compression, namely {\em linearization}, i.e. the transformation of the input matrix into a 1D string which is then compressed. We study the effectiveness of this technique for the simple row-by-row linearization and the more complex linearization based on the Peano-Hilbert space-filling curve;

\fullversion{
\item we show that the measures for 2D strings introduced in this paper can be generalized to $\Dim$-dimensional strings for any $\Dim>0$ with similar properties.
}

\end{itemize}

Overall our results shed some light on the difficulties of detecting and exploiting repetitiveness in the 2D setting, and show that some concepts/tools introduced in 1D are less effective in 2D.
Our results also show that to fully capture the repetitiveness in 2D strings it is necessary to consider the repetition also of non-square substrings; this suggest that to make the 2D Block Tree block effective it should consider also non-square partitions. 

Representations of 2D strings based on grammar compression (approaching the measures $g$ and $g_{rl}$) and macro schemes (approaching the measure~$b$) appear to be the most compact for some families of 2D strings, and those based on grammar compression also support 
efficient access to the symbols of the uncompressed string. Such results suggest that it may be worthwhile to study how to generalize the heuristics for building 1D grammars to 2D strings, and whether such heuristics maintains their ability to produce grammars of size provably close to the optimal~\cite{Bannai_Hirayama_Hucke_Inenaga_Jez_Lohrey_Reh_2021,NOP20}.

\section{Notation and background}\label{sec:notation}

Let $\Sigma = \{a_1, a_2, \ldots, a_\sigma\}$ be a finite ordered set of \emph{symbols}, which we call an \emph{alphabet}.
A \emph{2D string} $\MathM_{m \times n}$ is a ($m \times n$)-matrix with $m$ \emph{rows} and $n$ \emph{columns} such that each element $\MathM[i][j]$ belongs to $\Sigma$. The \emph{size} of $\Mmn$, denoted by $|\MathM|$, is $N = mn$. Note that a position in $\Mmn$ consists of a pair $(i,j)$, with $1\leq i \leq m$ and $1\leq j \leq n$. 
Throughout the paper, we assume that for each 2D string $\Mmn$ it holds that $m,n\geq 1$. Note that traditional one-dimensional strings are a special case of 2D strings with $m=1$.
We denote by $\Sigma^{m\times n}$ the set of all matrices with $m$ rows and $n$ columns over $\Sigma$. 
A 2D string in $\Sigma^{m\times n}$ is called \emph{square} if $m=n$. 

The concatenation between two matrices is a partial operation that can be performed horizontally ($\ohrz$) or vertically ($\ovrt$), with the constraint that the two operands must have the same number of rows (for $\ohrz$) or columns (for $\ovrt$).

\fullversion{
\begin{example}\label{ex:2D_concatenation}Consider the 2D strings 
$$A = \begin{bmatrix}
\syma & \symb & \syma\\
\syma & \symb & \symb
\end{bmatrix},\,  B = \begin{bmatrix}
\symb & \symb & \syma & \symb\\
\symb & \symb & \symb & \symb
\end{bmatrix},\text{ and } C = \begin{bmatrix}
\syma & \syma & \syma\\
\syma & \syma & \symb\\
\syma & \symb & \symb
\end{bmatrix}.$$
We can obtain new 2D strings by using $\ohrz$ and $\ovrt$, respectively:
$$A \ohrz B = \begin{bmatrix}
\syma & \symb & \syma & \symb & \symb & \syma & \symb\\
\syma & \symb & \symb & \symb & \symb & \symb & \symb
\end{bmatrix}, \text{ and } A \ovrt C = \begin{bmatrix}
\syma & \symb & \syma\\
\syma & \symb & \symb\\ 
\syma & \syma & \syma\\
\syma & \syma & \symb\\
\syma & \symb & \symb
\end{bmatrix}.$$
Note that $A \ovrt B$ and $A \ohrz C$ are undefined.
\end{example}
The concatenation operations have been described in~\cite{GiammarresiR97} where concepts and techniques of formal languages have been generalized to two dimensions. Such operations have been used in~\cite{BermanKLPR02} to define Straight-Line Programs for 2D strings that we will recall and generalize in Section~\ref{sec:slp}.
}

We denote by $\Mmn[i_1 \dd i_2][j_1 \dd j_2]$ the submatrix starting at position $(i_1, j_1)$ and ending at position $(i_2, j_2)$. We say that a matrix $F$ is a \emph{factor} or \emph{substring} of $\Mmn$ if there exist two positions $(i_1, j_1)$ and $(i_2, j_2)$ such that $F = \Mmn[i_1 \dd i_2][j_1 \dd j_2]$.
Given a 2D string $\Mmn$, the \emph{2D substring complexity} function $P_\MathM$ counts for each pair of positive integers $(k_1,k_2)$ the number of distinct ($k_1\times k_2$)-factors in $\Mmn$. 

\fullversion{
\begin{example}\label{ex:2D_factor}Consider the 2D strings $$M = \begin{bmatrix}
\syma & \syma & \symb & \symb\\
\syma & \syma & \symb & \symb\\
\syma & \syma & \symb & \symb\\
\syma & \syma & \symb & \symb\\
\syma & \syma & \symb & \symb
\end{bmatrix},\,F_1 =  \begin{bmatrix}
\syma & \symb \\
\syma & \symb  \\
\syma & \symb  \\
\end{bmatrix},\,F_2 =  \begin{bmatrix}
\syma & \syma \\
\syma & \syma 
\end{bmatrix},\,F_3 =  \begin{bmatrix}
\symb & \symb \\
\symb & \symb 
\end{bmatrix}\text{ and } F_4 = \begin{bmatrix}
\syma & \symb \\
\syma & \symb 
\end{bmatrix}.$$
The 2D string $F_1$ is a $(3\times 2)$-factor of $M$, as $F_1 = M[2\dd 4][2\dd 3]$. Moreover, it can be verified that $F_2, F_3$ and $F_4$ are the only $(2 \times 2)$-factors of $M$. Hence, $P_M(2,2) = 3$.
\end{example}
}

\fullversion{The main purpose of this paper is to generalize the repetitiveness measures $\delta$, $\gamma$, $b$, $g$, and $g_{rl}$ introduced in the last decade for 1D strings~\cite{NavarroSurvey} to 2D strings and higher dimensional strings. Some initial results in this area have been recently obtained in~\cite{CarfagnaManzini2024} with the definition of generalizations of the measures  $\delta$, $\gamma$, and $b$ to 2D strings looking at their {\em square} factors. Such measures will be recalled and analyzed in this paper using the symbols 
$\deltaCM$, $\gammaCM$, and $\bCM$. Working with only square factors ensures that $\deltaCM$ can be computed in linear time, and make this measure useful for the analysis of the 2D block tree by Brisaboa et al~\cite{BrisaboaGGN24}. In this paper we consider repetitiveness measures defined in terms also of non-square factors since we are interested in their expressive power regardless of algorithmic considerations, and we want to compare them with grammar-induced measures whose definitions involve non-square factors.}

\section{Measures \texorpdfstring{$\delta$}{Delta} and \texorpdfstring{$\gamma$}{Gamma} for 2D Strings}

In this section we extend to 2D strings the notion of $\delta$ measure~\cite{ChristiansenEKN21} and the notion of attractor~\cite{KP18} and its associated measure~$\gamma$.

\begin{definition} \label{def:2ddelta}Let $\Mmn$ be a 2D string and $P_\MathM$ be the 2D substring complexity of $\Mmn$. Then, $\delta(\Mmn) = \max\{P_\MathM(k_1, k_2)/k_1k_2, 1\leq k_1\leq m, 1\leq k_2\leq n\}$.
\end{definition}

Note that for 1D strings (i.e. $m=1$) the above definition coincides with the one used in the literature~\cite{KociumakaNP23,NavarroSurvey}. 
Recently, in~\cite{CarfagnaManzini2023} Carfagna and Manzini introduced an alternative extension of $\delta$, here denoted by $\delta_\square$, limited to square 2D input strings and using only {\em square} factors for computing the substring complexity. Below we report the definition of such a measure, applied to a generic two-dimensional string.

\begin{definition}
    Let $\Mmn$ be a 2D string and $P_\MathM$ be the 2D substring complexity of $\Mmn$. Then, $\delta_\square(\Mmn) = \max\{P_\MathM(k, k)/k^2, 1\leq k\leq \min\{m,n\}\}$.
\end{definition}

From the definitions of $\delta_\square$ and $\delta$, the following lemma easily follows.

\begin{lemma}
For every 2D string $\Mmn$ it holds that $\delta(\Mmn) \geq \delta_\square(\Mmn)$.
\end{lemma}

Although the two measures $\delta$ and $\delta_\square$ may seem similar, considering square factors instead of rectangular ones may result in very different values. Example~\ref{ex:deltasquare_on_strings} shows how different the two measures can be when applied to one-dimensional strings, while Example~\ref{ex:Carfagna_Manzini_Lemma4} shows that there exist families of square 2D strings for which $\delta_\square=o(\delta)$.

\begin{example}
\label{ex:deltasquare_on_strings}
    Given a 1D string $S\in\Sigma^n$, let $\MathM_{1\times n}\in\Sigma^{1\times n}$ be the matrix such that $\MathM_{1 \times n}[1][1\dd n] = S[1 \dd n]$. Since the only squares that occur in $\MathM_{1\times n}$ are the factors of size $1\times 1$, it is $\delta_\square(\MathM_{1\times n}) = P_\MathM(1,1)/1^2 \leq |\Sigma|$. On the other hand, $\delta(\MathM_{1\times n}) = \delta(S)$.
\end{example}

\begin{example}\label{ex:Carfagna_Manzini_Lemma4}
Let $\MathM_{n\times n}$ be the square 2D string in~\cite[Lemma 4]{CarfagnaManzini2023}. Assuming $n$ is an even perfect square, the first row of $\MathM_{n\times n}$ is the string $S= B_1B_2 \dd B_{\sqrt{n}/2}$ composed by $\sqrt{n}/2$ blocks, each one of size $2\sqrt{n}$, with $B_i = \one^i\zero^{(2\sqrt{n}-i)}$. The remaining rows of $\MathM_{n \times n}$ are all $\texttt{\#}^n$. In~\cite[Lemma 4]{CarfagnaManzini2023} it is shown that $\deltaCM(\MathM_{n \times n}) = O(1)$. On the other hand, notice that for $i \in [2\dd\sqrt{n}/2]$ and $j \in [0 \dd \sqrt{n}-i]$, the strings $\zero^j\one^i\zero^{\sqrt{n}-j-i}$ are all different substrings of length $\sqrt{n}$ of $S$. Since these substrings are in total $\Omega(n)$, it is $\delta(\MathM_{n \times n}) = \Omega(\sqrt{n})$.
\end{example}


\begin{definition}\label{def:2dstrinattractor}An  \emph{attractor} for a 2D string $\Mmn$ is a set $\Gamma \subseteq \mxn$ with the property that any substring $\MathM[i\dd j][k \dd l]$ of $\Mmn$ has an occurrence $\MathM[i'\dd j'][k' \dd l']$ such 
that $\exists (x, y) \in \Gamma$ with $i' \le x \le j'$ and $k' \le y \le l'$. The size of the smallest attractor for $\Mmn$ is denoted by $\gamma(\Mmn)$.
\end{definition}

When $m=1$ the above definition coincides with the one for 1D strings, hence the measure $\gamma$ inherits the properties for the one-dimensional case \cite{MantaciRRRS21,KP18}. In particular: $\gamma$ is not monotone and computing $\gamma(\Mmn)$ is NP-hard. In addition, the following property holds.

\begin{proposition}\label{prop:delta<=gamma}
For every 2D string $\Mmn$, it is $\delta(\Mmn)\leq \gamma(\Mmn)$.   
\end{proposition}
\begin{proof}
Reasoning as in the 1D case, observing that any attractor position belongs to at most $k_1 k_2$ factors of shape $k_1 \times k_2$, we get that for any 2D string $\MathM$ it is $P_\MathM(k_1,k_2)\leq k_1 k_2 \gamma(\MathM)$.  
\qed 
\end{proof}

The next proposition shows that in the 2D context, the gap between $\delta$ and $\gamma$ can be larger than the one-dimensional case, where it is logarithmic~\cite{KociumakaNP23}.

\begin{proposition}\label{prop:gamma_odelta} For all $m,n\geq1$ there exists a 2D string $\Mmn$ such that $\delta(\Mmn)=O(1)$ and $\gamma(\Mmn) = \Omega(\min(m,n))$.
\end{proposition}

\begin{figure}[t]
\centering
\begin{tikzpicture}
\matrix [matrix of math nodes,column sep=.5cm,row sep=.5cm] (m)
{
    \one & \zero & \zero & \zero & \zero & \zero  & {\underline{\zero}} & \zero  & \zero & \zero \\
    \zero &  \underline{\one} & \zero & \zero & \zero & \zero  & \zero & \zero  & \zero & \zero \\
    \zero & \zero & \underline{\one} & \zero & \zero & \zero  & \zero & \zero  & \zero & \zero \\
    \zero & \zero & \zero & \underline{\one} & \zero & \zero  & \zero & \zero  & \zero & \zero \\
    \zero & \zero & \zero & \zero & \underline{\one} & \zero  & \zero & \zero  & \zero & \zero \\
    \zero & \zero & \zero & \zero & \zero & \underline{\one}  & \zero & \zero  & \zero & \zero \\
    \underline{\zero} & \zero & \zero & \zero & \zero & \zero  & \one & \underline{\zero}  & \zero & \zero \\
};
\draw (m-1-1.north west) -- (m-1-6.north east) -- (m-1-6.south east) -- (m-1-1.south west) -- (m-1-1.north west);
\draw (m-2-2.north west) -- (m-2-7.north east) -- (m-2-7.south east) -- (m-2-2.south west) -- (m-2-2.north west);
\draw[dashed,->]  (m-1-1.west) to [bend right=45] (m-2-2.west);
\draw (-4.5,3.5) rectangle (4.5,-3.5);
\end{tikzpicture}
\caption[2D string attractor for $\Mmn$]{2D string attractor for the matrix $\Mmn$ of Proposition \ref{prop:gamma_odelta}. The cells whose positions belong to the string attractor are underlined. We show how $M[1 \dd 1][1 \dd 6]$ has an occurrence $M[2 \dd 2][2 \dd 7]$ crossing the string attractor position $(2,2)$.}
\label{fig:fig_example_2d_attractor}
\end{figure}
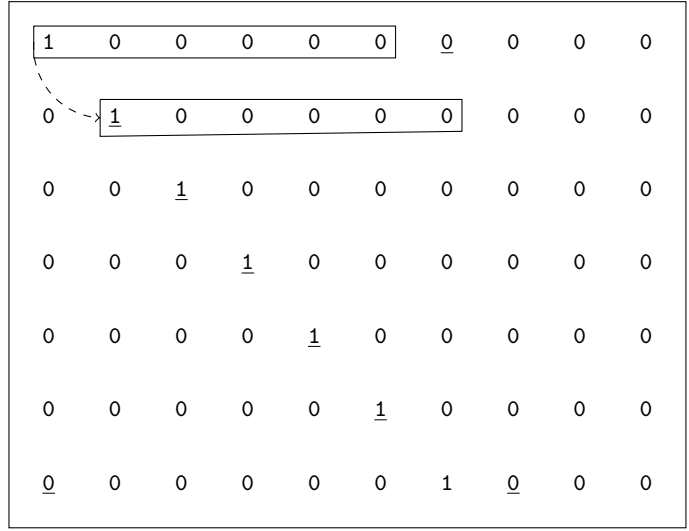

\begin{proof}
Let $I_k$ be the $k \times k$ identity matrix. For all $m, n\geq 1$, let us consider the 2D string $\MathM_{m\times n}$ such that $\MathM_{m \times n}[1\dd \min(m,n)][1\dd \min(m,n)] = I_{\min(m,n)}$, and all the remaining symbols are \zero's (see Fig.~\ref{fig:fig_example_2d_attractor} for an example). When either $m=1$ or $n=1$ the proof is trivial from known results on 1D strings, so let us assume $m,n\geq2$. Let us further assume $m<n$, next we show that $\Gamma = \{(2,2) \dd (m-1,m-1)\} \cup \{(1, m), (m, 1), (m, m+1)\}$ is an attractor for $\MathM_{m\times n}$. All the substrings of $\MathM_{m\times n}$ that contain at least two occurrences of $\one$'s have an occurrence crossing the position $(i, i)$, for some $1< i <m$, while all the substrings that consist of only $\zero$'s have an occurrence aligned with one of the $\zero$'s at position $(1,m)$, $(m,1)$, or $(m, m+1)$. The factors left contain only one occurrence of $\one$'s that do not cross any position in $\Gamma$. These factors of size $k_1 \times k_2$ have to cross either the $\one$ in position $(1,1)$ or in position $(m,m)$, and therefore it is either $k_1=1$ and $k_2<m$, or vice versa. Observe that all these factors have another occurrence either starting in $(2,2)$ or ending in $(m-1, m-1)$, and therefore $\Gamma$ is an attractor of $\MathM_{m\times n}$. Since $\MathM_{m\times n}$ has $m+1$ distinct columns the above attractor has minimum size, i.e. $\gamma(\MathM_{m\times n}) = m+1$. On the other hand, there exist at most $k_1+k_2$ distinct substrings of size $k_1 \times k_2$ in $\MathM_{m\times n}$: $k_1+k_2-1$ correspond to substrings where the diagonal of $\MathM_{m\times n}$ touches ones of the positions in the left or upper borders of the factor; the last one is the string of only $\zero$'s. Hence $\delta(\Mmn)\leq 2$. The case $m>n$ is treated symmetrically by considering the attractor $\Gamma' = \{(2,2) \dd (n-1,n-1)\} \cup \{(1, n), (n, 1), (n+1, n)\}$. For the case $n=m$ it is $\Mmn=I_n$ and reasoning as above it is easy to see that $\Gamma'' = \{(2,2) \dd (n-1,n-1)\} \cup \{(1, n), (n, 1)\}$ of size $n$ is a minimal attractor for $I_n$ and that $\delta(I_n)\leq 2$.
\qed 
\end{proof}

In~\cite{CarfagnaManzini2023} the authors introduced an alternative definition of string attractors for square 2D input strings in which they consider only square factors. We can define such a measure, denoted by $\gamma_\square$, also for generic 2D strings, by simply considering only square substrings of $\Mmn$ in Definition~\ref{def:2dstrinattractor}.
From the definitions of $\gamma$ and $\gamma_\square$ we immediately get the following relationship:

\begin{lemma}
  For every 2D string $\Mmn$ it holds that $\gamma(\Mmn) \geq \gamma_\square(\Mmn)$.  
\end{lemma}

The following example shows that $\gamma$ and $\gamma_\square$ can be asymptotically different.

\begin{example}\label{ex: gammaCMIdentity}
    Consider again the $m\times m$ identity matrix $I_m$. 
    For each $k\leq m$, a $k\times k$ square factor of $I_m$ either consists of i) all $\zero$'s, or ii) all $\zero$'s except only one diagonal composed by $\one$'s.
    Hence, all square factors of type i) have an occurrence that includes position $(m,1)$ (i.e. the bottom left corner), while all those of type ii) have an occurrence that includes the position $(\lfloor m/2 \rfloor, \lfloor m/2 \rfloor)$ (i.e. the $\one$ at the center).
    It follows that $\gamma_\square(I_m) = 2 \in O(1)$, while from the proof of Proposition~\ref{prop:gamma_odelta} it can be deduced that $\gamma(I_m) = \Theta(m)$. 
\end{example}

\fullversion{An important feature that many practical compressibility measures usually have is \emph{reachability}. A measure $\mu$ is reachable if we can represent any string in $O(\mu(w))$ space. We can generalize this notion to 2D strings.
\begin{definition}
    A 2D compressibility measure~$\mu$ is \emph{reachable} if every string $w$ can be represented in $O(\mu(w))$ space.
\end{definition}
The measures $\delta$ and $\gamma$ inherit from the 1D case the property that $\delta$ is unreachable and $\gamma$ is unknown to be reachable~\cite{NavarroSurvey}; but we will  introduce some reachable 2D measures in the following sections.}

\fullversion{One of the original motivations for considering the measure $\delta$ is that in the 1D case it can be computed in linear time and small extra space (see~\cite{bernardiniISAAC.2023} and references therein).
Since it is based only on square factors the measure~$\deltaCM$ can be still computed in linear time~\cite[Theorem~2]{CarfagnaManzini2024}, while no linear time algorithm is known for computing $\delta$ for 2D strings. However, by simply enumerating all factors $\delta(\Mmn)$ can be computed in time polynomial in $\max({m,n})$.}

\section{(Run-Length) Straight-Line Programs for 2D Strings}\label{sec:slp}

In this section, we consider a generalization of SLPs for the two-dimensional space introduced in \cite{BermanKLPR02} and use it to generalize the measures $g$ and $g_{rl}$ to 2D strings.

\begin{definition}
Let $\Mmn$ be a 2D string. A \emph{2-dimensional Straight-Line Program} (2D SLP) for $\Mmn$ is a context-free grammar $(V, \Sigma, R, S)$ that uniquely generates $\Mmn$ where $V$ is the set of non-terminal symbols or variables, $\Sigma$ is the set of terminal symbols, $S\in V$ is the axiom/starting symbol of the grammar and $R$ is the set of rules. A rule $A\rightarrow \alpha$ in $R$ with $A\in V$ can have one of the following three forms depending on its right-hand side $\alpha$: $$A \rightarrow a,\; A \rightarrow B\ohrz C,\; \text{ or } A  \rightarrow B \ovrt C$$ where $a \in \Sigma$, $B, C \in V$. We call these definitions \emph{terminal rules}, \emph{horizontal rules}, and \emph{vertical rules}, respectively and their corresponding expansion is defined as 
$$\gexp(A) = a,\quad \gexp(A) = \gexp(B)\ohrz \gexp(C),\; \text{ or } \gexp(A) = \gexp (B) \ovrt \gexp(C)$$

The size $|G|$ of a 2D SLP $G$ is the sum of the sizes of all the rules of $G$, where we assume that the terminal rules have size $1$, and the horizontal and vertical rules have size $2$. 
The measure $g(\Mmn)$ is defined as the size of the smallest 2D SLP generating $\Mmn$.
\end{definition}

\fullversion{Note that, by our definition of the $\ohrz$ operator, a horizontal rule $A \to B \ohrz C$ requires that the number of rows of $\gexp(B)$ and $\gexp(C)$ coincides, and the same must be true for the number of columns for a vertical rule $A \to B \ovrt C$.
We further assume that two distinct variables $A,B\in V$ do not have the same right-hand side.}

\fullversion{For 2D SLP it is convenient to introduce the concepts of {\em parse tree} and {\em grammar tree}, see Figures~\ref{fig:SLP} and~\ref{fig:prunedSLP} respectively.


\begin{definition}
    Let $G = (V,\Sigma, R,S)$ be a 2D SLP. 
    The \emph{parse tree} $T_G$ of $G$ is a directed labeled graph $T_G = T(S)$ obtained recursively as follows. For each variable $A\in V$:
    \begin{itemize}
        \item If $A \rightarrow a \in R$, then $T(A)$ is 
        a tree with root $A$ having a single child which is the leaf $a$.
        \item If $A \rightarrow B \ohrz C \in R$, then $T(A)$ is a tree with root $A$, $T(B)$ as its left subtree, and $T(C)$ as its right subtree. 
        \item If $A \rightarrow B \ovrt C \in R$, then $T(A)$ is a tree with root $A$, $T(B)$ as its up subtree, and $T(C)$ as its down subtree. 
    \end{itemize}

\end{definition}

Note that all the nodes with the same label are considered different. 
By \emph{traversal of the parse tree} $T_G$, we mean the usual visit in preorder of the parse tree of the 2D SLP $G$, with the following peculiarity: for each node $A$ corresponding to a horizontal rule $A \rightarrow B \ohrz C$, we first visit the node $A$, then the left subtree $T(B)$, and then the right subtree $T(C)$; for each node $A$ corresponding to a vertical rule $A \rightarrow B \ovrt C$, we first visit the node $A$, then the up subtree $T(B)$, and then the down subtree $T(C)$.

The tree $T_G$ has $\gexp(\Mmn) = N$ leaves. 
For simplicity, given a tree node labeled $A$, we use $\gexp(A)$ to denote the 2D string obtained by expanding the corresponding rule $A\rightarrow \alpha$ of $G$. A node in $T_G$ that has no braching and corresponds to a production rule of the form $A \rightarrow a$ (i.e. $\gexp(A)\in \Sigma$) is called \emph{leaf-generating node}.

\begin{definition}
\label{def:gtree}
Let $G = (V,\Sigma, R,S)$ be a 2D SLP.
We call \emph{primary occurrence} of a variable $A\in V$ the first node labeled with $A$ encountered in the traversal of the parse tree $T_G$. 
The remaining nodes labeled with the same $A$ are then called \emph{secondary occurrences}.
A \emph{grammar tree} for $G$ is constructed from the parse tree $T_G$ by keeping all the nodes that are either primary occurrences or children of a primary occurrence, while the children of every secondary occurrence are pruned. 
\end{definition}}




\ignore{
\begin{figure}
    \centering
    \scalebox{0.85}{\begin{minipage}[t]{.4\linewidth}
            \vspace{0pt}
            \begin{tikzpicture}[scale=.9]
            \draw (-0.5,3.5) rectangle (5.5,-0.5);
                \foreach \x in {0,1,...,5}{
                    \foreach \y in {0,1,...,3}{
                       \ifthenelse{\isodd{\x}}{\node at (\x,\y){\one};}{\node at (\x,\y){\zero};}
                }}
            \end{tikzpicture}
        \end{minipage}}    
    \caption{The $4\times 6$ binary 2D string used in our examples.}
    \label{fig:010101}
\end{figure}}

\fullversion{
\begin{figure}
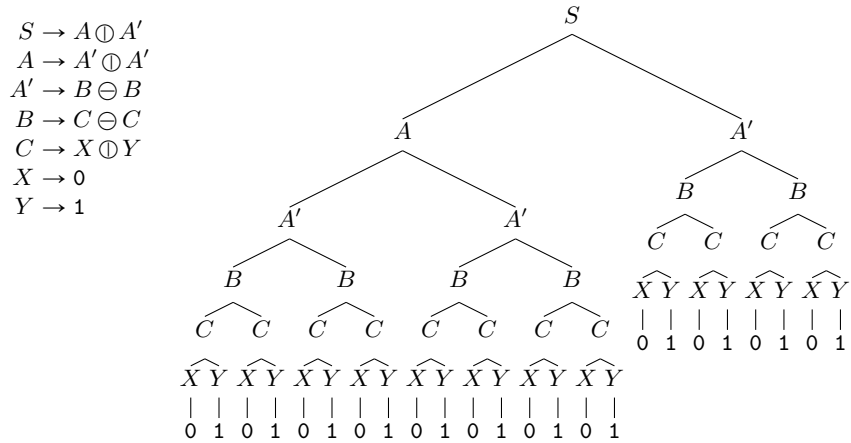

    \hspace*{\fill}
    \begin{minipage}[t]{\linewidth}
         \begin{minipage}[t]         {.2\linewidth}
         \vspace{0pt}
             \[ 
                  \begin{array}{rl}
                    S &\rightarrow  A\ohrz A' \\
                    A&\rightarrow A'\ohrz A' \\
                    A' &\rightarrow B\ovrt B\\ 
                    B&\rightarrow C\ovrt C\\
                    C&\rightarrow X \ohrz Y\\
                    X &\rightarrow \zero\\
                    Y & \rightarrow \one
                  \end{array}
             \] 
        \end{minipage}
        \begin{minipage}[t]{0.2\linewidth}
            \Tree [.$S$ [.$A$ [.$A\1$ [.$B$ [.$C$ [ {\zero} ].$X$ [ {\one} ].$Y$ ] [.$C$ [ {\zero} ].$X$ [ {\one} ].$Y$ ] ] [.$B$ [.$C$ [ {\zero} ].$X$ [ {\one} ].$Y$ ] [.$C$ [ {\zero} ].$X$ [ {\one} ].$Y$ ] ] ] [.$A\1$ [.$B$ [.$C$ [ {\zero} ].$X$ [ {\one} ].$Y$ ] [.$C$ [ {\zero} ].$X$ [ {\one} ].$Y$ ] ] [.$B$ [.$C$ [ {\zero} ].$X$ [ {\one} ].$Y$ ] [.$C$ [ {\zero} ].$X$ [ {\one} ].$Y$ ] ] ] ] [.$A\1$ [.$B$ [.$C$ [ {\zero} ].$X$ [ {\one} ].$Y$ ] [.$C$ [ {\zero} ].$X$ [ {\one} ].$Y$ ] ] [.$B$ [.$C$ [ {\zero} ].$X$ [ {\one} ].$Y$ ] [.$C$ [ {\zero} ].$X$ [ {\one} ].$Y$ ] ] ] ]
        \end{minipage}        
    \end{minipage}
    \caption{Example of a 2D SLP $G = (V,\Sigma,R,S)$, where $V=\{S,A,A',B,C,X,Y\}$, $\Sigma= \{\zero,\one\}$, and the set of rules $R$ is displayed on the left. The parse tree is displayed on the right. The 2D string $\MathM=\gexp(S)$ is displayed in Figure~\ref{fig:prunedSLP} (right). By exhaustive search, the grammar $G$ has minimum size for $\MathM$, hence $g(\MathM) = 12$.}
    \label{fig:SLP}
\end{figure}}

\fullversion{

\begin{figure}
    \hspace*{\fill}
    \begin{minipage}[t]{\linewidth}
         \begin{minipage}[t]{.2\linewidth}
         \vspace{0pt}
            \centering
             \[ 
                  \begin{array}{rl}
                    S &\rightarrow  A\ohrz A' \\
                    A&\rightarrow A'\ohrz A' \\
                    A' &\rightarrow B\ovrt B\\ 
                    B&\rightarrow C\ovrt C\\
                    C&\rightarrow X \ohrz Y\\
                    X &\rightarrow \zero\\
                    Y & \rightarrow \one
                  \end{array}
             \] 
        \end{minipage}\qquad
        \begin{minipage}[t]{0.2\linewidth}
            \Tree [.$S$ [.$A$ [.$A\1$ [.$B$ [.$C$ [ {\zero} ].$X$ [ {\one} ].$Y$ ] [  ].$C$ ] [  ].$B$ ] [  ].$A\1$ ] [  ].$A\1$ ]
        \end{minipage}\qquad\qquad
        \scalebox{0.9}{\begin{minipage}[t]{.4\linewidth}
            \vspace{0pt}
            \begin{tikzpicture}[scale=.9]
            \draw (-0.5,3.5) rectangle (5.5,-0.5);
            \draw (-0.5,3.5) rectangle (0.5,2.5);
            \draw (0.5,3.5) rectangle (1.5,2.5);
            \draw (-0.5,1.5) rectangle (1.5,-0.5);
            \draw (1.5,3.5) rectangle (3.5,-0.5);
                \foreach \x in {0,1,...,5}{
                    \foreach \y in {0,1,...,3}{
                         \ifthenelse{\isodd{\x}}{\node at (\x,\y){\one};}{\node at (\x,\y){\zero};}
                }}
            \node at (-0.3,3.3){$X$};
            \node at (0.7,3.3){$Y$};
            \node at (3.7,3.3){$A'$};
            \node at (-0.3,2.3){$C$};
            \node at (-0.3,1.3){$B$};
            \end{tikzpicture}
        \end{minipage}}        
    \end{minipage}
    \caption{The same 2D SLP as in Figure~\ref{fig:SLP}, but instead of the parse tree we show the {\em grammar tree} where each variable is expanded exactly once. On the right, we show the generated binary 2D string with highlighted the expansion of some of the variables.}
    \label{fig:prunedSLP}
\end{figure}
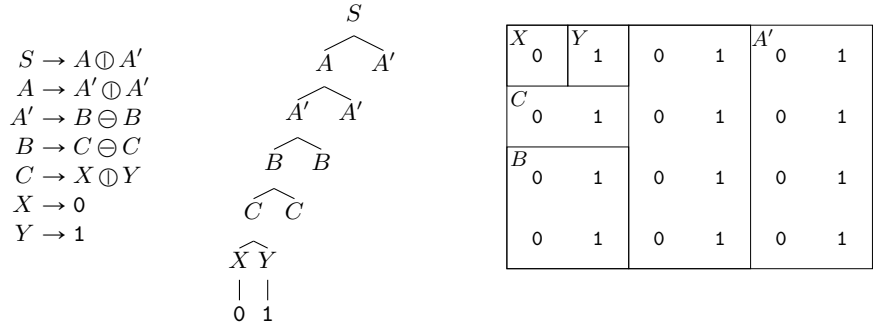 
}

\begin{proposition}\label{prop:lowerbound_g}
It always holds that $g(\Mmn) = \Omega(\log (mn))$. 
\end{proposition}

\begin{proof}
\fullversion{
The proof proceeds as the one of~\cite[Lemma~1]{Charikar2005} which proves the analogous result for strings and general 1D grammars. Given a matrix $M$ of size $N=nm$ and any 2D SLP $G$ for $M$, we build a sequence of non-terminals $\{A_i\}_{i=1,\ldots,k}$ defined as follows: $A_1$ is the starting symbol of $G$ and $A_{i+1}$ is a non-terminal symbol in the right-hand side $\alpha$ of the rule $A_i \rightarrow \alpha$ with maximal expansion. Note that since $G$ is acyclic all the $A_i$ must be distinct and therefore by continuing to extend the sequence we eventually reach a non-terminal $A_k$ with $|\exp(A_k)|=1$; moreover $k$ is at most the number of rules, hence $k\leq g$. Notice that for every rule $A_i\rightarrow \alpha$ the size of the expansion $|\exp(A_i)|$ is at most $2|\exp(A_{i+1})|$, that is, each rule application can at most double the size of the produced matrix. Therefore, it holds that $N=|\exp(A_1)|\leq 2|\exp(A_2)|\leq \cdots \leq 2^{k-1}|\exp(A_k)|<2^{g}$ and thus $g=\Omega(\log N)$.\qed}
\end{proof}

\begin{proposition}
The problem of determining if there exists a 2D SLP of size at most $k$ generating a text $\Mmn$ is NP-complete.
\end{proposition}

\begin{proof}
The problem belongs to NP because its 1D version, which is known to be NP-complete~\cite{Charikar2005}, can be reduced to the 2D version by interpreting 1D strings as matrices of size $1 \times n$. \qed
\end{proof}



As in the 1D case, we can extend 2D SLPs with \emph{run-length rules} obtaining more powerful grammars.

\begin{definition}
A \emph{2-dimensional Run-Length Straight-Line Program} (2D RLSLP) is a 2D SLP that in addition allows special rules, which are assumed to be of size 2, of the form
$$A \rightarrow  \ohrz^k B  \text{ and } A \rightarrow  \ovrt^k B$$
for $k > 1$, with their expansions defined respectively as 
\begin{align*}
    \gexp(A) &= \underbrace{\gexp(B)\ohrz \gexp(B) \ohrz \cdots \ohrz \gexp(B)}_{k \text{ times}} \\ \gexp(A) &=  \underbrace{\gexp(B)\ovrt \gexp(B) \ovrt \cdots \ovrt \gexp(B)}_{k \text{ times}}
\end{align*}
The measure $g_{rl}(\Mmn)$ is defined as the size of the smallest 2D RLSLP generating $\Mmn$.   
\end{definition}

\fullversion{

We also introduce the parse tree for 2D RLSLPs.

\begin{definition}
    Let a 2D RLSLP $G_{rl} = (V,\Sigma, R,S)$. 
    The \emph{parse tree} of $G_{rl}$ is a directed labeled graph $T_{G_{rl}} = T(S)$ obtained recursively as follows. For each variable $A$:
    \begin{itemize}
        \item If $A \rightarrow a \in R$, then $T(A)$ is 
        a tree with root $A$ having a single child which is the leaf $a$.
        \item If $A \rightarrow B \ohrz C \in R$, then $T(A)$ is a tree with root $A$, $T(B)$ as its left subtree, and $T(C)$ as its right subtree.
        \item If $A \rightarrow B \ovrt C \in R$, then $T(A)$ is a tree with root $A$, $T(B)$ as its up subtree, and $T(C)$ as its down subtree.
        \item If $A \rightarrow \ohrz^kB\in R$, then $T(A)$ is a tree with root $A$, with $k$ horizontal copies of $T(B)$ as its subtrees. 
        \item If $A \rightarrow \ovrt^kB\in R$, then $T(A)$ is a tree with root $A$, with $k$ vertical copies of $T(B)$ as its subtrees. 
    \end{itemize}

\end{definition}

As in the case of the parse tree of a 2D SLP, all the nodes with the same label are considered different. 
By \emph{traversal of the parse tree} $T_{G_{rl}}$, we extend the definition of the traversal of a parse tree of a 2D SLP as follows: for each node $A$ corresponding to a rule $A \rightarrow \ohrz^k B$, we first visit the node $A$, then each copy of the subtree $T(B)$ in order from the leftmost to the rightmost; for each node $A$ to a rule $A \rightarrow \ovrt^k B$, we first visit the node $A$, then each copy of the subtree $T(B)$ from the topmost to the bottommost.

From this notion of traversal of parse trees of 2D RLSLP's, the definition of grammar tree of a 2D RLSLP can be naturally derived from Definition~\ref{def:gtree}, with the exception that each node of the grammar tree associated to a run-length rule $A\rightarrow\ohrz^k B$ (resp. $A\rightarrow\ovrt^k B$) has two children as well: the left (resp. up) child is a node labeled by $B$, while the right (resp. down) child is a leaf labeled by $\ohrz^{k-1} B$ (resp. $\ovrt^{k-1} B)$, that is the remaining $k-1$ occurrences of $B$ collapse in a single node.
}

\begin{figure}
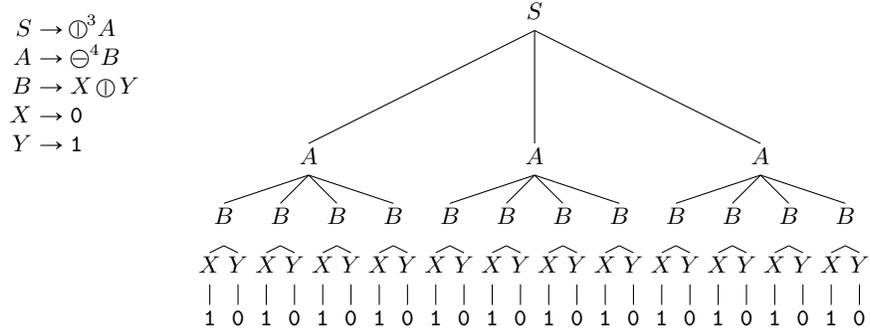

    \begin{minipage}[t]{\linewidth}
         \begin{minipage}[t]{.2\linewidth}
         \vspace{0pt}
             \[ 
                  \begin{array}{rl}
                    S &\rightarrow \ohrz^3 A \\
                    A&\rightarrow \ovrt^4 B \\
                    B&\rightarrow X\ohrz Y\\
                    X &\rightarrow \zero\\
                    Y & \rightarrow \one
                  \end{array} 
             \] 
        \end{minipage}
        \quad
        \begin{minipage}[t]{0.2\linewidth}
        \vspace{0pt}
            \Tree [.$S$  [.$A$ [.$B$ [ {\one} ].$X$ [ {\zero} ].$Y$ ] [.$B$ [ {\one} ].$X$ [ {\zero} ].$Y$ ] [.$B$ [ {\one} ].$X$ [ {\zero} ].$Y$ ] [.$B$ [ {\one} ].$X$ [ {\zero} ].$Y$ ] ] [.$A$ [.$B$ [ {\one} ].$X$ [ {\zero} ].$Y$ ] [.$B$ [ {\one} ].$X$ [ {\zero} ].$Y$ ] [.$B$ [ {\one} ].$X$ [ {\zero} ].$Y$ ] [.$B$ [ {\one} ].$X$ [ {\zero} ].$Y$ ] ] [.$A$ [.$B$ [ {\one} ].$X$ [ {\zero} ].$Y$ ] [.$B$ [ {\one} ].$X$ [ {\zero} ].$Y$ ] [.$B$ [ {\one} ].$X$ [ {\zero} ].$Y$ ] [.$B$ [ {\one} ].$X$ [ {\zero} ].$Y$ ] ] ]
        \end{minipage}
    \end{minipage}
    \caption{Example of a 2D RLSLP $G_{rl} = (V,\Sigma,R,S)$, where $V=\{S,A,B,X,Y\}$, $\Sigma= \{\zero,\one\}$, and the set of rules $R$ is displayed on the left. The parse tree is displayed on the right. The 2D string $\MathM=\gexp(S)$ is the depicted in Figure~\ref{fig:prunedSLP} (right). By exhaustive search, the grammar $G_{rl}$ has minimum size for $\MathM$, hence $g_{rl}(\MathM) = 8$.} 
    \label{fig:RLSLP}
\end{figure}

\begin{proposition} \label{prop:grl<=g}
For every 2D string $\Mmn$ it holds that  $g_{rl}(\Mnn) \leq g(\Mmn)$. Moreover, there are infinite string families where $g_{rl} = o(g)$. 
\end{proposition}
\fullversion{
\begin{proof} The first claim is trivial by definition. 
The second claim is proven by considering the family of $n\times n$ 2D strings $\MathM_{n\times n} = \zero^{n\times n}$: it is easy to see that the RLSLP $G_{rl} = (V, \Sigma, R, S)$, where $V =\{X,Y,Z\}$, $\Sigma = \{\zero\}$, $R= \{X\rightarrow\ovrt^nY, Y\rightarrow\ohrz^nZ, Z\rightarrow\zero\}$, and $S=X$, produces the string $\MathM_{n\times n}$, and therefore $g_{rl}=O(1)$. On the other hand, by Proposition~\ref{prop:lowerbound_g}, we have $g=\Omega(\log n)$, and the thesis follows.\qed
\end{proof}
}
\section{Efficient Direct Access on 2D (RL)SLPs}



\fullversion{
In this section, we show that a 2D RLSLP $G_{rl}$ representing a matrix of size $m \times n$ can be enriched with additional information to support random access to each element of the matrix in $O(\log mn)$ time and $O(|G_{rl}|\log mn)$ bits of space.
We first explain the result only for 2D SLPs, as it is easier to understand and generalize later to 2D RLSPLs. 
}

\fullversion{
Our approach builds upon the \emph{heavy-path decomposition} of Bille et al.~\cite{BilleLRSSW15}, defined for one-dimensional SLPs. In our setting, we consider the parse tree $T_G$ and we assign to each node labeled by non-terminal $A\in V$ the weight $|\gexp(A)|$. The \emph{heavy-path} associated to the node labeled by $A$ is the path $A_0, A_1, \ldots, A_k$ in 
$T_G$, where $A_i\in V$ for each $i=0,\ldots,k$, $A_0 = A$ and $\gexp(A_k)\in\Sigma$ (i.e. $A_k$ is the label of a leaf-generating node). Starting from $A$, at each step the heavy-path extends through the \emph{heaviest} child, i.e. the child with the greatest weight. In case of tie, the left child (in horizontal rules) or the up child (in vertical rules) is chosen, ensuring that each node is assigned a unique heavy path.  

The key idea behind our algorithm is as follows: for each non-terminal $A\in V$, we store the terminal symbol $a$ and its coordinates $(y_A, x_A)$ within $\gexp(A)$, which is reached by traversing the heavy-path in the parse tree. If the queried position $(y,x)$ coincides with $(y_A,x_A)$, we return $a$. Otherwise, by using efficient (opportune) data structures, we identify the lowest common ancestor (LCA) $A_i\in V$ of the leaves corresponding to the positions $(y_A, x_A)$ and $(y,x)$. We then recursively continue the search in the child $B$ of $A_i$ that was not included in the heavy-path of $A$, adjusting the query coordinates to $(y',x')\leftarrow(y-r_B+1,x-c_B+1)$ where $r_B$ and $c_B$ denote the top-left coordinates of the occurrence of $B$ within $A$. By starting the algorithm from the axiom $S$, it is easy to see that we will eventually access the element at the desired position. 
}


\fullversion{
This section is organized in the following way: we first describe how the nodes in a heavy-path are chosen from the parse tree of a 2D SLP; next, we describe the data structures that we need to efficiently perform lowest common ancestor queries; then, we show the space and time analysis of the algorithm; moreover, we describe the algorithm, and explain how to adapt the technique described to work with 2D RLSLP's; finally, we show how to optimize our data structures. 
}

\fullversion{
\subsection{2D heavy-path decomposition}
Let $G =(V,\Sigma, R, S)$ a 2D SLP generating $\Mmn$ and let $T_G$ be its parse tree. We consider the labels of the edges of $T_G$ connecting nodes labeled by non-terminal symbols. 
More formally, we denote by $E\subseteq V\times V\times\{\mathtt{L},\mathtt{R},\mathtt{U},\mathtt{D}\}$ the set of the \emph{edge labels} in the parse tree with leaves pruned, where each edge label is uniquely identified by the label of the parent, the label of the child and a spatial relationship as described below:
\begin{itemize}
    \item $(A,B,\mathtt{L}),(A,C,\mathtt{R})\in E$ if $A\rightarrow B\ohrz C\in R$;
    \item $(A,B,\mathtt{U}),(A,C,\mathtt{D})\in E$ if $A\rightarrow B\ovrt C\in R$.
\end{itemize}

}

\fullversion{
\begin{definition}
Let $G =(V, \Sigma, R, S)$ be a 2D 
SLP generating $\Mmn$ and $T_G$ its parse tree. We denote by $E_{\texttt{h}}\subseteq E$ the set of \emph{heavy edge labels} formed by: 

\begin{enumerate}
    \item $(A, B, \mathtt{L})$ if $A \rightarrow B \ohrz C \in R$ and $|\gexp(B)| \ge |\gexp(C)|$; 
    \item $(A, C, \mathtt{R})$ if $A \rightarrow B \ohrz C \in R$ and $|\gexp(B)| < |\gexp(C)|$;
    \item $(A, B, \mathtt{U})$ if $A \rightarrow B \ovrt C \in R$ and $|\gexp(B)| \ge |\gexp(C)|$;
    \item $(A, C, \mathtt{D})$ if $A \rightarrow B \ovrt C \in R$ and $|\gexp(B)| < |\gexp(C)|$.
\end{enumerate}
\end{definition}
\begin{definition}
The \emph{2D heavy-path decomposition} of $T_G$ is a directed labeled forest $H$ obtained by considering only the edges with label $e\in  E_\texttt{h}$. If a node $u$ labeled by $A$ has two children $v$ and $z$, labeled by $B$ and $C$ respectively, and $(A,B,p)\in E_h$, for some $p\in \{\Left,\Right, \Up, \Down\}$, then the node $v$ is called \emph{heavy child} of $u$, and $z$ is called \emph{light child} of $u$.
We call \emph{heavy-path} of a node $u$ any 
directed path in $H$ from $u$ to some leaf-generating node $v$. 
\end{definition}

We can associate each heavy path with the sequence of labels of its nodes. 
Note that the heavy paths of two nodes labeled with the same label are associated to the same sequence of labels. 

\begin{lemma}It holds $|E_{\texttt{h}}| = |V|- |\Sigma| = O(|G|)$.
\end{lemma}

A key property of the heavy edges is that not following them rapidly decreases the distance to a leaf of the parse tree.

\begin{lemma}\label{le:logn_heavye_edges}
Any path from the root to a leaf-generating node in $T_G$ traverses at most $\log_2 N$ edges not in $H$.
\end{lemma}
\begin{proof}
The result follows since traversing a light edge (i.e. an edge not in $H$) at least halves the size of the 2D string corresponding to the reached subtree, that is for a light edge $A\rightarrow B$ it is $|\gexp(A)|\geq 2|\gexp(B)|$.
\qed
\end{proof} 

}

\fullversion{

\subsection{The algorithm}\label{sec:alg}

Let $A_0, A_1,\ldots, A_k$ be the sequence of labels associated with a heavy-path.
The \emph{heavy symbol} of $A_i$, for all $i=0,\ldots, k$, is the symbol $\gexp(A_k)\in\Sigma$.
The \emph{heavy occurrence} of the symbol $\gexp(A_k)\in\Sigma$ within $\gexp(A_i)$ is defined recursively as the position $(y_i, x_i)$, where: 

$$y_i = \left\{
\begin{array}{lll}
1 & \text{if } i = k \\
y_{i+1} & \text{if } A_i \text{ corresponds to a horizontal rule or } A_i\rightarrow A_{i+1}\ovrt B\\
y_{i+1} + m &\text{if } A_i\rightarrow B\ovrt A_{i+1} \text{ with } \gexp(B)\in\Sigma^{m\times n}\\
\end{array}\right.$$

and

$$x_i = \left\{
\begin{array}{lll}
1 & \text{if } i = k \\
x_{i+1} & \text{if } A_i \text{ corresponds to a vertical rule or } A_i\rightarrow A_{i+1}\ohrz B\\
x_{i+1} + n &\text{if } A_i\rightarrow B\ohrz A_{i+1} \text{ with } \gexp(B)\in\Sigma^{m\times n}\\
\end{array}\right .$$
In other words, the heavy occurrence is the position in $\gexp(A_i)$ reached by following the heavy-path starting from $A_i$.

Suppose we are trying to access the position $(y,x)$ in $A_0$, i.e. $\gexp(A_0)[y][x]$.
As described at the beginning of this section, the algorithm checks whether the heavy occurrence $(y_0, x_0)$ of $\gexp(A_k)$ within $\gexp(A_0)$ is exactly the position we are looking for: if the positions coincide we simply return the letter explicitly stored; otherwise, we look in the heavy path for the lowest common ancestor $A_i$ between the leaves corresponding to the positions $(y,x)$ and $(y_0,x_0)$; then, we repeat recursively the procedure on the light child $B$ of $A_i$, this time though looking for the position $(y',x')$ that corresponds to the position $(y,x)$ moved by an offset derived from the top-left corner of this occurrence of $B$ in $A_0$, thus ensuring $\gexp(A_0)[y][x] = \gexp(B)[y'][x']$.

To retrieve the lowest common ancestor $A_i$ and to compute the updated position $(y',x')$, we need the following information for each heavy-path: 
\begin{itemize}
    \item the \emph{up size sequence} $u_0,u_1,\dots, u_k$,  where $u_i$ is the sum of the number of rows of the light up children of each of the first $i$ nodes in the heavy-path;
    
    \item the \emph{down size sequence} $d_0,d_1,\dots, d_k$,  where $d_i$ is the sum of the number of rows of the light down children of the first $i$ nodes in the heavy-path;
    
    \item the \emph{left size sequence} $l_0,l_1,\dots, l_k$,  where $l_i$ is the sum of the number of columns of the light left children of the first $i$ nodes in the heavy-pathl;
    
    \item the \emph{right size sequence} $r_0,r_1,\dots, r_k$,  where $r_i$ is the sum of the number of columns of the light right children of the first $i$ nodes in the heavy-path.
\end{itemize}

Each sequence above counts for each $i$ the number of rows above/below and columns to the left/right of the 2D substring $\gexp(A_i)$ within $\gexp(A_0)\in\Sigma^{m\times n}$ reached by following the heavy paths, that is, $$\gexp(A_0)[u_i+1\dd m-d_i][l_i+1\dd n-r_i] = \gexp(A_i).$$

The following lemma is a direct consequence \sloppy of the definitions of up/down/left/right size sequences.

\begin{lemma}\label{u<m-d,l<n-r}
    Let $A_0, A_1, \ldots, A_k$ be the sequence of labels associated with a heavy-path in a parse tree.
    The following relationships hold: \begin{itemize}
        \item  $u_0+1\leq u_1+1 \leq \ldots \leq u_k+1 = m-d_k \leq \ldots \leq m-d_0$;\item $l_0+1\leq l_1+1 \leq \ldots \leq l_k+1 = n-r_k \leq \ldots \leq n-r_0$.   \end{itemize}
\end{lemma}

To locate the lowest common ancestor between the leaves corresponding to the positions $(y_0,x_0)$ and $(y,x)$ in $\gexp(A_0)$, we look for the largest index $i$ such that \begin{equation}\label{eqy}
u_i+1\leq y\leq m - d_i
\end{equation} 
 and 
 \begin{equation}\label{eqx}
     l_i + 1 \leq x \leq n-r_i.
 \end{equation} 

Based on the values of $y$ with respect to $y_0$ we behave as follows: 

\begin{enumerate}
    \item[1.a] if $y>y_0$, we find the largest $i$ verifying $y\leq m-d_i$ by performing a predecessor query of $m-y+1$ in the down size sequence;
    \item[1.b] if $y<y_0$, we find the largest $i$ verifying $y\geq u_i+1$ by performing a predecessor query of $y$ in the up size sequence;
    \item[1.c] if $y=y_0$, return $i=k$. 
\end{enumerate}

We proceed analogously according to the values of $x$ with respect to $x_0$:
\begin{enumerate}
    \item[2.a] if $x>x_0$, we find the largest $j$ verifying $x\leq n-r_j$ by performing a predecessor query of $n-x+1$ in the right size sequence;
    \item[2.b] if $x<x_0$, we find the largest $j$ verifying $x\geq l_j+1$ by performing a predecessor query of $x$ in the left size sequence;    \item[2.c] if $x=x_0$, return $j=k$.
\end{enumerate}

Observe that if the conditions in~1.c and~2.c are both true, then we are done and we just return $\gexp(A_0)[y_0][x_0]$.
Otherwise, observe that by Lemma~\ref{u<m-d,l<n-r} only $i'=\min(i,j)$ verifies Equations~\ref{eqy} and~\ref{eqx} at the same time. 

Finally, we can recall the same procedure to the light child $B$ of $A_{i'}$, after updating the position $(y',x')$ to search within $\gexp(B)$. 
If we denote by $m_{i'+1}$ and $n_{i'+1}$ the number of rows and columns respectively in $A_{i'+1}$, we compute $(y',x')$ as follows:

$$y' = \left\{
\begin{array}{lll}
y - u_{i'}& \text{if } A_{i'} \text{ corresponds to a horizontal rule or } A_{i'}\rightarrow B\ovrt A_{i'+1}\\
y - u_{i'} - m_{i'+1} &\text{if } A_{i'}\rightarrow A_{i'+1}\ovrt B\\
\end{array}\right.$$

and

$$x' = \left\{
\begin{array}{lll}
x - l_{i'}& \text{if } A_{i'} \text{ corresponds to a vertical rule or } A_{i'}\rightarrow B\ohrz A_{i'+1}\\
x  - l_{i'} - n_{i'+1} &\text{if } A_{i'}\rightarrow A_{i'+1}\ohrz B\\
\end{array}\right.$$

}

\begin{figure}[ht]
\centering
\begin{tikzpicture}[scale=0.5]
    \draw[very thick] (0,0) rectangle (12, 12) [fill=gray, fill opacity=0.15]; 
    \draw[very thick] (1, 1) rectangle (6,10) [fill=gray, fill opacity=0.45]; 
    \draw[very thick] (6, 1) rectangle (10,10) [fill=gray, fill opacity=0.75]; 

    \draw[fill=black] (1.5,9) circle (0.1) node[above right] {$(y_0,x_0)$};
    \draw[fill=white] (8,3) circle (0.1) node[above right] {$(y,x)$};

    \draw (8,6) coordinate (0.1) node[above] {\Large $A_{i+1}'$};
    \draw (3.5,6) coordinate (0.1) node[above] {\Large $A_{i+1}$};
    
    \draw[thick, dashed, <->] (0.1,5.5) -- (0.9,5.5) node [midway, above] {$l_i$}; 
    \draw[thick, dashed, <->] (6.1,4.5) -- (11.9,4.5) node [midway, above] {$r_{i+1}$};
    \draw[thick, dashed, <->] (10.1,5.5) -- (11.9,5.5) node [midway, above] {$r_i$};
    \draw[thick, dashed, <->] (3.5,10.1) -- (3.5,11.9) node [midway,right] {$u_i$};
    \draw[thick, dashed, <->] (3.5,0.1) -- (3.5,0.9) node [midway,right] {$d_i$};

    \draw[thick, dotted, <->] (0.1,9) -- (1.3,9) node [midway, above] {$l_k$}; 
    \draw[thick, dotted, <->] (1.7,9) -- (11.9,9) node [midway, above] {$r_k$};
    \draw[thick, dotted, <->] (1.5,9.2) -- (1.5,11.9) node [midway,right] {$u_k$};
    \draw[thick, dotted, <->] (1.5,0.1) -- (1.5,8.8) node [midway,right] {$d_k$};

    \draw[thick, <->] (1.1,3) -- (5.9,3) node [midway,above] {$n_{i+1}$};
\end{tikzpicture}
\caption{Components of a heavy-path labeled  $A_0,\dots,A_k$. The 2D string $\gexp(A_0)$ is represented by the biggest rectangle.
The black dot at coordinate $(y_0,x_0)$ is the heavy-occurrence of $\gexp(A_k)$ in $\gexp(A_0)$.
The rectangle in light gray is $\gexp(A_{i+1})$, and in dark gray is $\gexp(A_{i+1}')$. They are generated by the rule $A_i \rightarrow A_{i+1} \ohrz A'_{i+1}$. $A_{i+1}$ and $A'_{i+1}$ are the heavy child and light child of $A_i$, respectively.
We further show the values of the up/down/left/right sequence corresponding to the variables $A_i$, $A_{i+1}$, and $A_k$. For $A_{i+1}$ we only represent $r_{i+1}$ since $r_{i+1}\neq r_i$, while $l_{i+1}=l_i$, $u_{i+1}=u_i$ and $d_{i+1}=d_i$.
The white dot at coordinates $(y,x)$ is $\gexp(A_0)[y][x]$, the cell we want to access. Note that $A_i$ is the lowest common ancestor between the heavy-path endpoint $\gexp(A_0)[y_0][x_0]$ and the leaf $\gexp(A_0)[y][x]$. After finding $A_i$, the search must continue within $\gexp(A_{i+1}')$. More precisely, we look for the position $(y - u_i, x - (l_i+n_{i+1}))$ within $\gexp(A_{i+1}')$.} 
\label{fig:heavy-paths}
\end{figure}
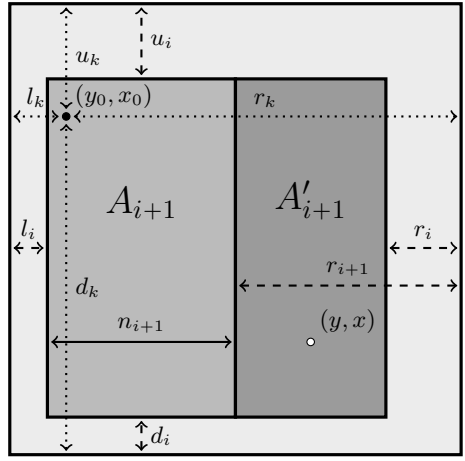

Figure \ref{fig:heavy-paths} illustrates heavy-paths and direct access on 2D SLPs.

\fullversion{
\subsection{Data structures}\label{sec:data_structures}


Recall that the heavy-paths of two distinct nodes with the same non-terminal symbol share the same sequence of labels. This means that the sequence of labels $A_{0_j}, A_{1_j}, \ldots A_{k_j}$ associated with a heavy path of a node labeled by $V_j=A_{0_j}\in V$ is uniquely determined by $V_j$. Then, for each variable $V_j\in V$, with $j=1,\ldots, |V|$ we need to store the following information:

\begin{itemize}
\item The number of rows $m_{0_j}$ and columns $n_{0_j}$ of the 2D string $\gexp(A_{0_j})$.
\item The indexes $y_{0_j}$ and $x_{0_j}$ of the \emph{heavy occurrence} of  $\gexp(A_{k_j})$ within $\gexp(A_{0_j})$, and the symbol $\gexp(A_{k_j})[1][1] = \gexp(A_{0_j})[y_{0_j}][x_{0_j}]$. 
\item The \emph{up size sequence} $u_{0_j},u_{1_j},\dots, u_{k_j}$;
\item The \emph{down size sequence} $d_{0_j},d_{1_j},\dots, d_{k_j}$;

\item The \emph{left size sequence} $l_{0_j},l_{1_j},\dots, l_{k_j}$;
\item The \emph{right size sequence} $r_{0_j},r_{1_j},\dots, r_{k_j}$.
\end{itemize}

Note that the values $m_{0_j}, n_{0_j}, x_{0_j}, y_{0_j}$, and $\gexp(A_{k_j})$ depend only on $V_j$ itself, hence we can store them for all the variables using $5|V|$ words of space. 

To compactly represent all heavy paths and efficiently support predecessor queries, we construct a data structure that encodes the up, down, left and right sequences. This structure is an adaptation to the two-dimensional context of the heavy-path suffix forest defined in~\cite{BilleLRSSW15}, which allows us to reduce a predecessor query to a \emph{weighted ancestor query}~\cite{predecessor_FM}. We consider the forest $F$ of tries obtained by reversing the direction of the edges of the heavy-path decomposition $H$ of the parse tree $T_G$. In this new representation, nodes that were previously leaf-generating become the roots of the new forest $F$. This means that each trie in $F$ corresponds to a distinct terminal symbol from the alphabet $\Sigma$ (produced by a leaf-generating node). Then, the number of tries in $F$ is at most $\Sigma$. Moreover, every non-terminal symbol appears exactly once in the entire forest $F$. The core of our data structure is the weighted forest $F$ that stores, for each edge labeled $(A_{j_{i+1}}, A_{j_{i}}, p)$, where $p\in \{\mathtt{L}, \mathtt{R}, \mathtt{U}, \mathtt{D}\}$, four spatial weights to encode the contribution of light children: a \emph{left weight} $w_l$, a \emph{right weight} $w_r$, an \emph{up weight} $w_u$, and a \emph{down weight} $w_d$, defined as follows:
\begin{itemize}
    \item if $p=\mathtt{L}$ (i.e. $A_i\rightarrow A_{i+1}\ohrz B$), then $w_l=w_u=w_d=0$ and $w_r = n_B$;
    \item if $p=\mathtt{R}$ (i.e. $A_i\rightarrow B\ohrz A_{i+1}$), then $w_r=w_r=w_u=0$ and $w_l = n_B$;
    \item if $p=\mathtt{U}$ (i.e. $A_i\rightarrow A_{i+1}\ovrt B$), then $w_l=w_r=w_u=0$ and $w_d = m_B$;
    \item if $p=\mathtt{D}$ (i.e. $A_i\rightarrow B\ovrt A_{i+1}$), then $w_l=w_r=w_d=0$ and $w_u = m_B$, 
\end{itemize}  
where $n_B$ and $m_B$ denote the number of columns and the number of rows in the matrix $\gexp(B)$, respectively.
Thus, if a node labeled $B$ is a light child of a node labeled $A_i$ in the parse tree $T_G$, these values represent the contribution of the light child $B$ to the spatial weights.
This data structure integrates both the heavy-path decomposition and the weighted forest, enabling efficient retrieval of lowest common ancestors and fast computation of updated positions, as described in the next subsection. By using the formulation of Bille et al.~\cite{BilleLRSSW15}, performing a weighted ancestor query on the up/down/left/right weights is equivalent to performing a predecessor query on the up/down/left/right size sequence, respectively.
Each query is performed in $O(\log \log \max (m,n))$ time~\cite{predecessor_FM}, if $G$ generates the 2D string $\Mmn$.

}


\fullversion{
\subsection{Space and time complexities}\label{sec:space_time}

By Lemma \ref{le:logn_heavye_edges}, there are at most $\log mn$ light edges in any path from root to a leaf-generating node in the parse tree of a 2D SLP $G$ generating $\Mmn$. That is, we can change from one heavy-path to another at most $\log mn$ times. 
The data structures described in the previous paragraph take $O(|V|\log mn)=O(|G|\log mn)$ bits of space, can be built in $O(|G|)$ time, and support predecessor queries in $O(\log\log \max(m,n))$ time~\cite{predecessor_FM}.
Therefore, we obtain the following result.

\begin{theorem}\label{thm:direct_accessG}Let $\Mmn$ be a 2D string and let $G$ be a 2D SLP generating $\Mmn$.
We can build in $O(|G|)$ preprocessing time a data structure of $O(|G|\log mn)$ bits of space which supports direct access queries
to any cell $\MathM[i][j]$ in \sloppy
$O(\log mn \log \log \max(m,n))$ time.
\end{theorem}
}

\fullversion{
\subsection{Generalization to 2D RLSLPs}

The strategy based on heavy-paths can be generalized to work on 2D RLSLPs, by adding specific additional considerations related to the run-length rules. In particular, given the heavy-path $A_0, A_1, \ldots, A_k$: 
\begin{enumerate}
    \item when a node $u$ labeled by $A_i$ corresponds to the rule $A_i\rightarrow\ohrz^\ell A_{i+1}$ (resp. $A_i\rightarrow\ovrt^\ell A_{i+1}$), the heavy child of $u$ is the leftmost (resp. upmost) child labeled by $A_{i+1}$; the remaining children are considered light;
    \item  in the definitions of heavy occurrence we add the conditions that if $A_i\rightarrow \ohrz^\ell A_{i+1}$ or $A_i\rightarrow \ovrt^\ell A_{i+1}$, then $y_i = y_{i+1}$ and $x_i = x_{i+1}$ 
    \item in the definitions of left and up size sequences, if $A_i\rightarrow \ohrz^\ell A_{i+1}$ or $A_i\rightarrow \ovrt^\ell A_{i+1}$, then $l_{i+1} = l_i$ and $u_{i+1} = u_{i}$; in the definitions of right (resp. down) size sequences, if $A_i\rightarrow \ohrz^\ell A_{i+1}$ (resp. $A_i\rightarrow \ovrt^\ell A_{i+1}$) and $\gexp(A_{i+1})\in\Sigma^{m_{i+1}\times n_{i+1}}$, then $r_{i+1} = r_{i} + (\ell-1)n_{i+1}$ (resp. $r_{i+1} = r_{i}$) and $u_{i+1} =u_i$ (resp. $u_{i+1}= u_{i} + (\ell-1)m_{i+1}$);
    \item let us denote by $i'$ the value returned when steps (1) and (2) described in Subsection~\ref{sec:alg} are applied to find the lowest common ancestor $A_{i'}$ between the leaves corresponding to the heavy symbol of $A_0$ and $\gexp(A_0)[y][x]$. If $A_{i'}\rightarrow \ohrz^\ell A_{i'+1}$ (resp. $A_{i'}\rightarrow \ovrt^\ell A_{i'+1}$) then $y' = 1+(y' - u_{i'}) \mod m_{i'+1}$ (resp. $y' = y' - u_{i'}$) and $x' = 1+(x' - l_{i'}) \mod n_{i'+1}$ (resp. $x' = x' - l_{i'}$).    
\end{enumerate}

Note that the considerations described above allow us to use the same algorithmic strategy presented in Subsection~\ref{sec:alg} and the same data structures defined in Subsection~\ref{sec:data_structures}, while preserving the same time and space complexities. So, we derive the following result.

\begin{theorem}\label{thm:direct_accessGRL}Let $\Mmn$ be a 2D string and let $G_{rl}$ be a 2D RLSLP generating $\Mmn$.
We can build in $O(|G_{rl}|)$ preprocessing time a data structure of $O(|G_{rl}|\log mn)$ bits of space which supports direct access queries
to any cell $\MathM[i][j]$ in 
$O(\log mn \log \log \max(m,n))$ time.
\end{theorem}

\fullversion{
\subsection{Speeding up the algorithm to $O(\log mn)$}

The time complexity described in Subsection~\ref{sec:space_time} is determined by the \sloppy $O(\log \log \max(m,n))$ time required for each of the $O(\log mm)$ weighted ancestor queries, where $mn$ is the size of the 2D string $\Mmn$.

Instead of performing both horizontal and vertical queries at each step of the algorithm, we can first execute only one of the two queries and, in $O(1)$ time, check whether the retrieved node is the lowest common ancestor $v$ between the heavy-occurrence $(y',x')$ of the heavy-path and the queried position $(y,x)$. If the first query does not succeed, we proceed with the second query, which is then guaranteed to return the correct lowest common ancestor $v$. However, in the best-case scenario, this optimization only reduces the number of queries by half, thus reaching asymptotically the same upper bound.


To get rid of the $O(\log \log \max(m,n))$ factor, we adapt the predecessor data structure, called \emph{interval-biased search tree}, proposed by Bille et al.~\cite{BilleLRSSW15} to our setting.  
A predecessor query $p$ on this data structure takes $O(\log U/U')$ time, where $U$ is the universe size and $U'$ is the gap between the successor and the predecessor of $p$. 
We use this data structure to obtain the following theorem.

\begin{theorem}
    Let $\Mmn$ be a 2D string and let $G_{rl}$ be a 2D RLSLP generating $\Mmn$. There exists a data structure that uses $O(|G_{rl}|\log mn)$ bits of space and supports direct access queries to any cell $\MathM[i][j]$ in $O(\log mn)$ time.
\end{theorem}
\begin{proof}
    We follow the same strategy described in Subsection~\ref{sec:alg}, with the 
key difference that we execute the queries on the left/right size sequence and on the up/down size sequence in parallel using the interval-biased search trees. As soon as we locate the node in the parse tree where the search should proceed (the light child of the lowest common ancestor), we terminate all other ongoing queries (if any). Note that each potential solution can be checked in $O(1)$ time, ensuring that the time complexity depends only on the query that successfully locates the lowest common ancestor.   
    
    We distinguish the query times to locate the lowest common ancestor (LCA) into two groups based on the corresponding type of rule: 
    \begin{enumerate}
    \item LCAs with up/down children. If we perform $l$ queries of this type, the complexity for $i$th query is $O(\log Y_i/Y'_{i})$, where $Y_i$ is the number of rows of the 2D string generated by the first node of the $i$-th heavy path, and $Y'_{i}$ is the number of rows of the 2D string generated by the light child of the LCA found in the query  
    \item LCAs with left/right children. If we perform $t$ queries of this type, the complexity for $j$th query is $O(\log X_j/X'_{j})$, where $X_j$ is the number of columns of the 2D string generated by the first node of the $j$-th heavy path, and $X'_{j}$ is the number of columns of the 2D string generated by the light child of the LCA found in the query.
    \end{enumerate}
    
    The total access time is then $$O\left(\sum_{i=1}^{l}\log Y_i/Y'_{i} + \sum_{j=1}^{t}\log X_j/X'_{j}\right).$$

    Since each LCA expansion represents a 2D substring of the expansion of the previous LCA found, the following inequalities hold: 
    $$n\geq X_1 \geq X'_1 \geq X_2 \geq X'_2\geq \ldots \geq X_{t}\geq X'_{t}\geq 1,$$ and $$m\geq Y_1\geq Y'_1 \geq Y_2 \geq Y'_2 \geq \ldots \geq Y_{l}\geq Y'_{l} \geq 1,$$
    where we assume $m$ and $n$ the number of rows and the number of columns of $\Mmn$, respectively. For simplicity, in the following equalities $m$ and $n$ are also denoted as $Y'_0$ and $X'_0$, respectively. 
    
    We obtain the final time 
    \begin{align*}
        O\left(\sum_{i=1}^{l}\log Y_i/Y'_{i} + \sum_{j=1}^{t}\log X_j/X'_{j}\right) = & O\left(\sum_{i=1}^{l}\log Y'_{i-1}/Y'_{i} + \sum_{j=1}^{t}\log X'_{j-1}/X'_{j}\right)\\
        =&O\left(\log m + \log n\right)\\
        =&O(\log mn).
    \end{align*}\qed 
\end{proof}

Therefore, the following corollary can be inferred. 

\begin{corollary}\label{thm:direct_access_grl}Let $\Mmn$ be a 2D string.
There exists a data structure that uses $O(g_{rl}\log mn)$  bits of space that supports direct access queries to any cell \sloppy $\MathM[i][j]$ in $O(\log  mn)$ time. 
\end{corollary}}
}

\section{Macro Schemes for 2D Strings}\label{sec:macroscheme}

The notion of macro scheme and the corresponding measure $b$ can be naturally generalized to 2D strings with the following definition.

\begin{definition}\label{def:macro}
A \emph{2D macro scheme} for a string $\Mmn$ is any factorization of $\Mmn$ into a set of disjoint phrases such that any phrase is either a square of dimension $1 \times 1$ called an \emph{explicit symbol/phrase}, or is a copied phrase with source in $\Mmn$ starting at a different position. For a 2D macro scheme to be  \emph{valid} or \emph{decodable}, the function $\map: ([1 \dd m] \times [1\dd n])  \cup \{\bot\} \rightarrow ([1 \dd m] \times [1\dd n]) \cup \{\bot\}$ induced by the factorization must verify that:


\begin{enumerate} 
    \item[i)] $\map(\bot) = \bot$, and if $\MathM[i][j]$ is an explicit symbol, then $\map(i,j) = \bot$;
    \item[ii)] for each copied phrase $\MathM[i_1\dd j_1][i_2\dd j_2]$, it must hold that $\map(i_1+t_1, i_2 + t_2) = \map(i_1, i_2) + (t_1, t_2)$ for $(t_1, t_2) \in [0\dd j_1-i_1] \times [0\dd j_2 - i_2]$, where $\map(i_1, i_2)$ is the upper left corner of the source for $\MathM[i_1\dd j_1][i_2 \dd j_2]$;
    \item[iii)] for each $(i, j) \in [1\dd m] \times [1 \dd n]$ there exists $k > 0$ such that $\map^k(i, j) = \bot$.
\end{enumerate}
We define $b(\Mmn)$ as the size of a smallest valid 2D macro scheme for $\Mmn$.
\end{definition}

\begin{example}
\label{ex:b_O1}Let $I_{n}$ be the $n \times n$ identity matrix. A macro scheme for $I_{n}$ consists of the phrases $\{X_1, X_2, X_3, X_4, X_5, X_6\}$ where: i) $X_1 = I_n[1][1]$ is an explicit symbol (the $\one$ in the top-left corner); ii) $X_2 = I_n[1][2]$ is an explicit symbol; $X_3 = I_n[2][1]$ is an explicit symbol; $X_4 = I_n[1][3\dd n]$ is a phrase with source $(1,2)$; $X_5 = I_n[3\dd n][1]$ is a phrase with source $(2,1)$; and $X_6 = I_n[2 \dd n][2\dd n]$ is a phrase with source $(1,1)$. The underlying function $\map$ is defined as $\map(1,1) = \map(1, 2) = \map(2,1) = \bot$, $\map(1,j) = (1, j-1)$ for $j \in [3\dd n]$, $\map(i, 1) = (i-1, 1)$ for $i \in [3\dd n]$, and $\map(i,j) = (i-1, j-1)$ for $i, j \in [2\dd n] \times [2\dd n]$. One can see that $\map^n(i,j) = \bot$ for each $i$ and $j$. Hence, the macro scheme is valid and $b(I_n) \le 6$. Figure~\ref{fig:fig1} shows this macro scheme for~$I_7$.
\end{example}

\begin{figure}[t]
\centering
\begin{tikzpicture}[scale=0.75]
    [
        box/.style={rectangle,draw=black,thick, minimum size=1cm},
    ]

\foreach \x in {0,1,...,6}{
    \foreach \y in {0,-1,...,-6}{
         \ifthenelse{\x=-\y}{\node at (\x,\y){\one};}{\node at (\x,\y){\zero};}
}}
\draw (-0.5,0.5) rectangle (0.5,-0.5);
\draw (0.5,-0.5) rectangle (6.5,-6.5);
\draw (-0.5,0.5) rectangle (6.5,-6.5);
\draw (-0.5,-1.5) rectangle (0.5,-1.5);
\draw (0.5,0.5) rectangle (1.5,-0.5);
\draw[dashed, ->] (0.8,-0.8) to[bend right] (0.2,-0.2);
\draw[dashed, ->] (0,-1.7) to[bend right] (0,-1.3);
\draw[dashed, ->] (1.7,0) to[bend right] (1.3, 0);
\end{tikzpicture}
\caption{Macro scheme with 6 phrases for $I_7$. The entries 
$(1,1)$, $(1,2)$, and $(2,1)$ are explicit symbols.  The remaining phrases point to the source from where they are copied.}
\label{fig:fig1}
\end{figure}

The following two propositions show that the computability properties of $b$ and its relationship with the measures $g_{rl}$ and $g$ are preserved in the 2D context.

\begin{proposition}
The problem of determining if there exists a valid 2D macro scheme of size at most $k$ for a text $\Mmn$ is NP-complete.
\end{proposition}

\begin{proof}
The 1D version of the problem, which is known to be NP-complete~\cite{Gallant1982}, reduces to the 2D version of the problem in constant time. \qed
\end{proof}

\begin{proposition}
\label{prop:b<=grl}
For every 2D string $\Mmn$ it holds that $b(\Mmn) \leq g_{rl}(\Mmn)$.
\end{proposition}
\fullversion{
\begin{proof}
We show how to construct a macro scheme from a 2D RLSLP, representing the same 2D string and having the same asymptotic size. Let $G_{rl}$ be a 2D RLSLP generating $\Mmn$ and consider its grammar tree. Each leaf of the grammar tree corresponding to a variable that expands to a single symbol at cell $\Mmn[i][j]$, induces an explicit phrase of the parsing at that specific cell. A leaf of the grammar tree corresponding to an occurrence of the variable $A$ expanding at cells $\Mmn[i_1\dd i_2][j_1\dd j_2]$ becomes a phrase of the parsing at these cells, and its source is aligned with the upper-left corner of the expansion on $\Mmn$ of the primary occurrence of $A$ in the grammar tree. \sloppy For a leaf $\ohrz^{k-1}B$ induced by an horizontal run-length variable, which expands to $\Mmn[i_1\dd i_2][j_1\dd j_2]$ in $\Mmn$, we construct a phrase at these cells, pointing to the occurrence of $\gexp(B)$ at its left in $\Mmn$. We do analogously, for 2D vertical run-length rules.  This parsing is decodable and its size is bounded by the size $g_{rl}$ of the grammar.\qed
\end{proof}}


\fullversion{
\begin{figure}
    \centering
    \begin{minipage}[t]{.40\linewidth}
            \vspace{0pt}
            \begin{tikzpicture}{scale=.7}
                \draw (-0.5,3.5) rectangle (0.5,2.5);
                \draw (0.5,3.5) rectangle (1.5,2.5);
                \draw (1.5,3.5) rectangle (5.5,-0.5);
                \draw (-0.5,2.5) rectangle (1.5,-0.5);
                \draw[dashed, ->] (1.6,3.4) to[bend right] (0,3.4);
                \draw[dashed, ->] (-0.2,2.2) to[bend left] (-0.2,2.8);
                \foreach \x in {0,1,...,5}{
                    \foreach \y in {0,1,...,3}{
                         \ifthenelse{\isodd{\x}}{\node at (\x,\y){\one};}{\node at (\x,\y){\zero};}
                }}
            \end{tikzpicture}
        \end{minipage} 
    \caption{Example of 2D macro scheme derived from the 2D RLSLP in Figure~\ref{fig:RLSLP}.}
    \label{fig:b_from_grl}
\end{figure}

In the case of 1D strings it is already known that for the Fibonacci words it holds $b = o(g_{rl})$ \cite{latin/GagieNP18,NOP20}. The following result shows that the same result holds even for square 2D matrices.

\begin{proposition}\label{prop: g_rlIdentity}
There exists an infinite family of square 2D strings of size $N$ where $g_{rl} = \Omega(b\log N)$.
\end{proposition}

\begin{proof}
Consider the identity matrix $I_n$ of order $n$ and a run-length 2D SLP $G$ for $I_n$ of size $g$. In the following we call run-length symbols the non-terminals of $G$ corresponding to a run-length rule, and regular symbols the remaining non-terminals. We preliminarily prove that $G$ can generate the $1$s of $I_n$ only by expanding regular symbols: consider without loss of generality an horizontal run-length rule $A \rightarrow \ohrz^k B$ of $G$ with $k>1$, if $\exp(A)$ contains at least a $1$ then the same must hold for $\exp(B)$ and therefore $\exp(A)$, and thus $I_n$, would contain at least $k>1$ $1$s in the same row which is impossible. 

Given a non-terminal symbol $A$, we denote with $\hash A$ the number of $1$s in $\exp(A)$. Similarly to what we have done in the proof of Proposition \ref{prop:lowerbound_g} we consider the sequence of non-terminals $\{A_i\}_{i=1\ldots k}$ defined as follows: $A_1$ is the starting symbol of $G$ and $A_{i+1}$ is a non-terminal symbol in the right-hand side $\alpha$ of the rule $A_i \rightarrow \alpha$ which maximizes the number of $1$s in its expansion. Since $\exp(A_1)=I_n$ it is $\hash A_1= n$ and $A_1$ cannot be a run-length symbol, therefore it is $\hash A_1 \leq 2 \hash A_2$ and $A_2$ cannot be a run-length symbol otherwise it would be $\hash A_2=0$ and $\hash A_1 = 0$. By iterating this reasoning we inductively build a chain $n=\hash A_1 \leq 2 \hash A_2 \leq \cdots \leq 2^{k-1}\hash A_k$ of inequalities in which no run-length symbol appears. Similarly to what we observed in the proof of Proposition \ref{prop:lowerbound_g} it holds that $A_k$ corresponds to a terminal rule $A_k\rightarrow 1$ with $k\leq g$. As a consequence, $n \leq 2^{k-1} < 2^{g}$ and therefore $g=\Omega(\log n)$. The results follows since $\log n = \Theta(\log N)$ and $b(I_n)=O(1)$ (see Example~\ref{ex:b_O1}).\qed
\end{proof}
}

In~\cite{CarfagnaManzini2024} the authors introduce a notion of 2D macro scheme for square strings that differs from the one in Definition~\ref{def:macro} in that 1) phrases must all be square and 2) phrases are allowed to overlap (see~\cite[Section~4]{CarfagnaManzini2024} for details). 
Given a 2D string $\MathM$, in the following we call $\bCM(\MathM)$ the minimum number of phrases in a macro scheme with square, possibly overlapping, phrases.  
We now show that the use of square phrases can limit significantly the power of a macro scheme since there are string families for which $b = o(\bCM)$. 
This result will have important consequences also for other measures. 

\begin{figure}
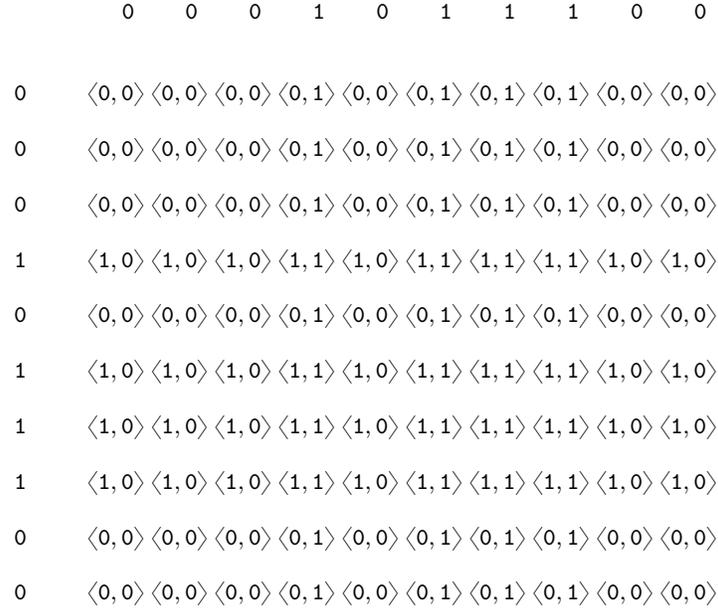

    \centering
    \begin{tabular}{p{25pt} c c c c c c c c c c}
            & \quad \zero & \quad \zero & \quad \zero &\quad  \one &\quad  \zero & \quad \one & \quad \one & \quad \one & \quad \zero & \quad \zero\\[20pt]
    \zero   & $\langle\zero,\zero\rangle$ & $\langle\zero,\zero\rangle$ & $\langle\zero,\zero\rangle$ & $\langle\zero,\one\rangle$ & $\langle\zero,\zero\rangle$ & $\langle\zero,\one\rangle$ & $\langle\zero,\one\rangle$ & $\langle\zero,\one\rangle$ & $\langle\zero,\zero\rangle$ & $\langle\zero,\zero\rangle$\\[10pt]
    \zero   & $\langle\zero,\zero\rangle$ & $\langle\zero,\zero\rangle$ & $\langle\zero,\zero\rangle$ & $\langle\zero,\one\rangle$ & $\langle\zero,\zero\rangle$ & $\langle\zero,\one\rangle$ & $\langle\zero,\one\rangle$ & $\langle\zero,\one\rangle$ & $\langle\zero,\zero\rangle$ & $\langle\zero,\zero\rangle$\\[10pt]
    \zero   & $\langle\zero,\zero\rangle$ & $\langle\zero,\zero\rangle$ & $\langle\zero,\zero\rangle$ & $\langle\zero,\one\rangle$ & $\langle\zero,\zero\rangle$ & $\langle\zero,\one\rangle$ & $\langle\zero,\one\rangle$ & $\langle\zero,\one\rangle$ & $\langle\zero,\zero\rangle$ & $\langle\zero,\zero\rangle$\\[10pt]
    \one   & $\langle\one,\zero\rangle$ & $\langle\one,\zero\rangle$ & $\langle\one,\zero\rangle$ & $\langle\one,\one\rangle$ & $\langle\one,\zero\rangle$ & $\langle\one,\one\rangle$ & $\langle\one,\one\rangle$ & $\langle\one,\one\rangle$ & $\langle\one,\zero\rangle$ & $\langle\one,\zero\rangle$\\[10pt]
    \zero   & $\langle\zero,\zero\rangle$ & $\langle\zero,\zero\rangle$ & $\langle\zero,\zero\rangle$ & $\langle\zero,\one\rangle$ & $\langle\zero,\zero\rangle$ & $\langle\zero,\one\rangle$ & $\langle\zero,\one\rangle$ & $\langle\zero,\one\rangle$ & $\langle\zero,\zero\rangle$ & $\langle\zero,\zero\rangle$\\[10pt]
    \one   & $\langle\one,\zero\rangle$ & $\langle\one,\zero\rangle$ & $\langle\one,\zero\rangle$ & $\langle\one,\one\rangle$ & $\langle\one,\zero\rangle$ & $\langle\one,\one\rangle$ & $\langle\one,\one\rangle$ & $\langle\one,\one\rangle$ & $\langle\one,\zero\rangle$ & $\langle\one,\zero\rangle$\\[10pt]
    \one   & $\langle\one,\zero\rangle$ & $\langle\one,\zero\rangle$ & $\langle\one,\zero\rangle$ & $\langle\one,\one\rangle$ & $\langle\one,\zero\rangle$ & $\langle\one,\one\rangle$ & $\langle\one,\one\rangle$ & $\langle\one,\one\rangle$ & $\langle\one,\zero\rangle$ & $\langle\one,\zero\rangle$\\[10pt]
    \one   & $\langle\one,\zero\rangle$ & $\langle\one,\zero\rangle$ & $\langle\one,\zero\rangle$ & $\langle\one,\one\rangle$ & $\langle\one,\zero\rangle$ & $\langle\one,\one\rangle$ & $\langle\one,\one\rangle$ & $\langle\one,\one\rangle$ & $\langle\one,\zero\rangle$ & $\langle\one,\zero\rangle$\\[10pt]
    \zero   & $\langle\zero,\zero\rangle$ & $\langle\zero,\zero\rangle$ & $\langle\zero,\zero\rangle$ & $\langle\zero,\one\rangle$ & $\langle\zero,\zero\rangle$ & $\langle\zero,\one\rangle$ & $\langle\zero,\one\rangle$ & $\langle\zero,\one\rangle$ & $\langle\zero,\zero\rangle$ & $\langle\zero,\zero\rangle$\\[10pt]
    \zero   & $\langle\zero,\zero\rangle$ & $\langle\zero,\zero\rangle$ & $\langle\zero,\zero\rangle$ & $\langle\zero,\one\rangle$ & $\langle\zero,\zero\rangle$ & $\langle\zero,\one\rangle$ & $\langle\zero,\one\rangle$ & $\langle\zero,\one\rangle$ & $\langle\zero,\zero\rangle$ & $\langle\zero,\zero\rangle$
    \end{tabular}
    \caption{Example of 2D string $B_3$ on the alphabet $\{\langle\zero,\zero\rangle,\langle\zero,\one\rangle,\langle\one,\zero\rangle,\langle\one,\one\rangle\}$. It is based on the de Bruijn word $D_3 = \zero \zero \zero \one \zero \one \one \one \zero \zero$, displayed horizontally and vertically respectively above and on the left of $B_3$.}
    \label{fig:2DdeBuijn}
\end{figure}

For any $k>0$, let $D_k$ denote a binary de Bruijn sequence of length $n=2^k + k -1$ containing all the possible binary substrings of length $k$ exactly once. We define the $n \times n $ matrix $B_k$ over the alphabet $\Delta = \{\langle \zero, \zero \rangle,\langle \zero, \one \rangle,\langle \one, \zero \rangle,\langle \one, \one \rangle\}$ by the relationship
$$
B_k[i][j] = \langle D_k[i], D_k[j] \rangle.
$$
Notice that each row and each column of $B_k$ is a de Bruijn sequence over a binary alphabet which is a subset of $\Delta$. For example, if $D_k[i] = \one$, then row~$i$ of $B_k$ is a de Bruijn sequence over the alphabet $\{\langle \one, \zero \rangle,\langle \one, \one \rangle\}$. Similarly, if $D_k[j]=\zero$, then column $j$ of $B_k$ is a de Bruijn sequence over the alphabet $\{\langle \zero, \zero \rangle,\langle \one, \zero \rangle\}$.
Notice that $B_k$ contains only two distinct rows/columns since rows/columns $i$ and $j$ are different if and only if $D_k[i]\neq D_k[j]$.
\fullversion{An example is shown in Figure~\ref{fig:2DdeBuijn}.}

\begin{lemma} \label{lemma:gBk}
For any $k>0$ it is $g(B_k) = O(n\log\log n/\log n)$, where $B_k$ has size $n\times n$, with $n=2^k+k-1$.
\end{lemma}

\begin{proof}
   In~\cite{latin/GagieNP18} the authors show that $g(D_k) = O(n\log\log n/\log n)$, that is, there exists a (one dimensional) SLP $G$ of size $O(n\log\log n/\log n)$ generating the de Bruijn sequence $D_k$. If in $G$ we replace the terminal symbols $\{ \zero, \one\}$ with  $\{\langle \zero, \zero \rangle,\langle \zero, \one \rangle\}$ we obtain a SLP $G_0$ that generates all rows $i$ in $B_k$ such that $D_k[i]=\zero$. Similarly, if in $G$ we replace the terminals $\{ \zero, \one\}$ with  $\{\langle \one, \zero \rangle,\langle \one, \one \rangle\}$ we obtain a SLP $G_1$ that generates all rows $i$ in $B_k$ such that $D_k[i]=\one$. Finally, if in $G$ we make all rules vertical and we replace the terminal $\zero$ with the starting symbol of $G_0$ and the terminal $\one$ with the starting symbol of $G_1$, we obtain a 2D SLP $G_2$ that combined with $G_0$ and $G_1$ generates the matrix $B_k$. The thesis follows since the size of $G_2 \cup G_1 \cup G_0$ is $O(n\log\log n/\log n)$. \qed
\end{proof}

\begin{lemma}\label{lemma:nok}
    Every square submatrix of size $k$ or more appears in $B_k$ at most once. 
\end{lemma}

\begin{proof}
    It suffices to prove the result for the square submatrices of size $k$. Assume that the $k \times k$ submatrix with top-left corner in $(i,j)$ is identical to the submatrix with top-left corner in $(u,v)$. The crucial observation is that $B_k[i][j] = B_k[u][v]$ implies $D_k[i] = D_k[u]$ and  $D_k[j] = D_k[v]$. Considering also the other entries, we get that if the two submatrices are equal then we must have $D_k[i..i+k-1] = D_k[u..u+k-1]$ and $D_k[j..j+k-1] = D_k[v..v+k-1]$. Since $D_k$ is a de Bruijn sequence, this implies $i=u$ and $j=v$ as claimed.  \qed
\end{proof}

\begin{proposition}\label{prop:bvsbsq}
    For the 2D string $B_k$ of size $N = n\times n$, with $n=2^k+k-1$, it is $\bCM = \Omega(g \sqrt{N}/(\log N\log\log N))  = \Omega(b \sqrt{N}/(\log N\log\log N))$. 
\end{proposition}

\begin{proof}
    Since $b\leq g$, by Lemma~\ref{lemma:gBk} $b=O(\sqrt{N}\log\log N/\log N)$. Lemma~\ref{lemma:nok} implies that there cannot be square phrases of size $k\times k$ or larger. Hence, the number of phrases is at least $n^2/k^2$, so $\bCM = \Omega(N/\log^2 N)$ and the lemma follows. \qed
\end{proof}

\section{On the Relative Power of 2D Measures} \label{sec:differences}

A remarkable property of the measures considered in this paper is that, for 1D strings, they can be totally ordered in terms of their relative power at capturing regularities in the input; indeed, for any 1D string $S$ it is $\delta(S) \leq \gamma(S) \leq b(S) \leq g_{rl}(S) \leq g(S)$  (see \cite{KP18,NavarroSurvey}).
In the previous sections, we have shown that when we consider also 2D strings the relationships $\delta\leq\gamma$ and $b\leq g_{rl}\leq g$ still hold. In this section, however, we prove that the two classes of measures, i.e. $\delta$ and $\gamma$ from one side, based on the counting and distribution of distinct factors, and $b$, $g_{rl}$, and $g$ from the other side, based on copy-paste mechanisms, become incomparable when also 2D strings are considered. In particular $g$ can be asymptotically smaller than $\delta$:

\begin{proposition}\label{prop:g_odelta}
There exists an infinite family of 2D strings of size $N$ with $\delta = \Omega(g N/\log^3 N)$. 
\end{proposition}
\begin{proof}
For $k\geq1$, consider the 2D binary string $E_k$ of size $N=k \times 2^k$ such that, for all $1\leq i \leq 2^k$, the $i$th column of $E_k$ is the binary representation of $i-1$ in $k$ digits (with the top row containing the least significant bits). Since all columns of $E_k$ are distinct, $E_k$ contains $2^k$ distinct factors of size $k \times 1$ and therefore it is $\delta(E_k) \geq 2^k/k=kN/k^3$. We prove that $g(E_k)=O(k)$ by exhibiting a 2D SLP for~$E_k$ of size $O(k)$ and the result immediately follows because $k<\log N$.

Consider the 2D SLP $G_k = (V_k, \{\zero, \one\}, R'_k, S_{k})$ having the following rules $R'_k$:
\begin{itemize}
    \item $X_0 \rightarrow \zero$ and $X_h \rightarrow X_{h-1} \ohrz X_{h-1}$ for all $1 \leq h \leq k-1$;
    \item $Y_0 \rightarrow \one$ and $Y_h \rightarrow Y_{h-1} \ohrz Y_{h-1}$ for all $1 \leq h \leq k-1$;
    \item $C_h \rightarrow X_{h-1} \ohrz Y_{h-1}$ for all $2 \leq h \leq k$;
    \item $S_1 = X_0 \ohrz Y_0$ and $S_h \rightarrow  R_{h} \ovrt C_{h}$ for all $2 \leq h \leq k$.
    \item $R_h \rightarrow S_{h-1} \ohrz S_{h-1}$ for all $2 \leq h \leq k$;
\end{itemize}
$G_k$ has size $\Theta(k)$ and it is easy to see that $\gexp(X_h)=\zero^{(2^h)}$, $\gexp(Y_h)=\one^{(2^h)}$ and therefore $\gexp(C_h)=\zero^{(2^{h-1})}\one^{(2^{h-1})}$. In the following, we show by induction on $k$ that $G_k$ is a 2D SLP for~$E_k$, that is $\gexp(S_k)=E_k$. 
For the base case $k=1$ it is $\gexp(S_1) = \gexp(X_0) \ohrz \gexp(Y_0) = \zero \ohrz \one=E_1$. For the inductive step we assume that $\gexp(S_k)=E_k$ and we note that for $k\geq 1$ the bottom row $E_{k+1}[k+1][1..2^{k+1}]$ of $E_{k+1}$ is the string $\zero^{(2^k)}\one^{(2^k)}=\gexp(C_{k+1})$ and the remaining rectangular submatrix $E_{k+1}[1..k][1..2^{k+1}]$ is $E_k \ohrz E_k$. 

By taking two expansion steps starting from $S_{k+1}$, we obtain $\gexp(S_{k+1})=(\gexp(S_k)\ \ohrz\ \gexp(S_k))\ \ovrt\ \gexp(C_{k+1})$ which by inductive hypotheses and the definition of $C_{k+1}$ expands to $(E_k \ohrz E_k) \ovrt E_{k+1}[k+1][1..2^{k+1}] = E_{k+1}$.
\qed
\end{proof}

From the above proposition, we get that the measure $g$ (and therefore also $g_{rl}$ and $b$) can be much smaller than both $\delta$ and $\gamma$. Intuitively, the reason is that the matrix $E_k$ is hard to compress by columns (they are all distinct) but easily compressible by rows. 
The measure $g$ is defined in terms of the best grammar compressor, so it fully exploits row compressibility. In contrast, the measure $\delta$ is based on occurrences of factors of any shape and is therefore ``hindered'' by the difficulty of compressing columns.
For 1D strings it is always $\delta \leq g$, but in the 2D setting, because of the greater freedom in choosing the shape of the copied patterns, the measures $b,\ g_{rl}$ and $g$ become mutually incomparable with both $\gamma$ and $\delta$. 

Given the above observation, it is worthwhile to compare $g$ with the measures $\deltaCM$ and $\gammaCM$ whose definitions are based on square factors. In some sense, using square factors can be seen as a method to capture both horizontal and vertical compressibility. Unfortunately, the following proposition shows that $g$ can be asymptotically smaller than $\deltaCM$.


\fullversion{
\begin{proposition}\label{prop:g_odeltaCM}
There exists a family of rectangular 2D strings of size $N$ with $\deltaCM = \Omega(g N/\log^4 N)$. 
\end{proposition}
\begin{proof}
Consider again the binary matrix $E_k$ of size $N=k\times 2^k$ of Proposition~\ref{prop:g_odelta} having $g(E_k)=O(k)$. We note that any $k \times k$ submatrix of $E_k$ is a distinct factor of size $k^2$ and therefore it is $\deltaCM(E_k)\geq (2^k-k+1)/k^2=\Omega(kN/k^4)$. The result immediately follows since $g(E_k)=O(k)$ and $k<\log N$. \qed
\end{proof}

The previous two propositions are based on rectangular matrices with a number of columns that is exponentially larger than the number of rows. Since measure $\deltaCM$ was originally proposed only for square input matrices, one may wonder whether the gap is due to the highly skewed shape of the input matrix. The next result shows that this is not the case in the sense that for input square matrices we have a smaller, but still significant, gap between $g$ and $\deltaCM$. Note that the same result also holds for $\gammaCM$, $\delta$, and $\gamma$, since $\deltaCM \leq \gammaCM$ and $\deltaCM \leq \delta \leq \gamma$.
}

\begin{proposition}\label{prop:g_odeltaCM_square}
There exists an infinite family of square 2D strings of size $N$ with $\deltaCM = \Omega(g \sqrt{N}/(\log N\log\log N))$.
\end{proposition}

\begin{proof}
    Consider the matrix $B_k$ defined in Section \ref{sec:macroscheme}. 
    By Lemma~\ref{lemma:gBk} it is $g(B_k) = O(n\log\log n/\log n) = O(\sqrt{N}\log\log N/\log N)$. 
    By Lemma~\ref{lemma:nok} $B_k$ contains $\Theta(n^2) = \Theta(N)$ distinct $k\times k$ factors, hence  $\deltaCM = \Omega(N/\log^2 N)$ and the bound follows. \qed
\end{proof}

Note that Propositions~\ref{prop:bvsbsq}, \ref{prop:g_odelta} and~\ref{prop:g_odeltaCM_square} show that in the 2D setting there can be a significant gap between $b$, $g_{rl}$ and $g$ from one side, and $\bCM$, $\deltaCM$, $\gammaCM$, $\delta$, $\gamma$ from the other. Furthermore, since the square matrix $\zero_{n\times n}$ consisting of only zeros has constant measures $\bCM$, $\deltaCM$, $\gammaCM$ but it is $g(\zero_{n\times n})=\Omega(\log n)$ (see Proposition \ref{prop:lowerbound_g}), Propositions~\ref{prop:bvsbsq} and \ref{prop:g_odeltaCM_square} imply that $g$ is also mutually incomparable with each of the above square measures, even if we consider only square input matrices. \fullversion{Moreover $g_{rl}$ is incomparable with both $\deltaCM$ and $\gammaCM$ since for the identity matrix $I_n$ it is $\deltaCM\leq\gammaCM=O(1)$ by Example~\ref{ex: gammaCMIdentity} and $g_{rl}=\Omega(\log n)$ by Proposition~\ref{prop: g_rlIdentity}.} 
 
We now show that even the recently introduced 2D Block Tree data structure~\cite{BrisaboaGGN24}, which is also based on square factors, can fail to capture the regularity of certain two-dimensional strings. The 2D Block Tree is a tree-like compressed representation of a square matrix supporting random access to individual entries in logarithmic time. Given an ${n\times n}$ input matrix $\MathM$, an integer parameter $c>1$, and assuming  that $n$ is a power of $c$, the root of the 2D Block tree at level $\ell=0$ represents the whole matrix $\MathM$. To build the level $\ell\geq 1$ of the tree we recursively partition (some of) the submatrices represented at level $\ell -1$ into $c^2$ smaller non-overlapping submatrices of size $n/c^\ell \times n/c^\ell$ called \emph{blocks}; for each of these blocks, the tree stores a corresponding descending node at level $\ell$. The 2D Block Tree attempts to compress the input matrix by avoiding the storage of redundant submatrices: if a block has a prior occurrence in row-major order (RMO), its corresponding subtree is candidate to be pruned and replaced with $O(1)$ pointers to the nodes overlapping its first occurrence in RMO. The pruned blocks are not partitioned into smaller matrices, and their corresponding nodes are leaves in the 2D block tree. See~\cite{BrisaboaGGN24,CarfagnaManzini2024} for details.

Unfortunately, the following example exhibits a family of 2D strings that are significantly more compressible when represented as an SLP compared to their 2D Block Tree representation: for these matrices, the 2D Block Tree fails to achieve a compression close to the measure $g$ (and therefore to $g_{rl}$ and $b$).

\begin{proposition}
There exists an infinite family of square 2D strings of size $N$ such that the number of nodes of their 2D Block Tree is $\Omega(g \sqrt{N}/(\log N\log\log N))$. The same result holds also for the attractor-based 2D Block Tree defined in~\cite[Theorem~5]{CarfagnaManzini2024}.
\end{proposition}

\fullversion{\begin{proof}    
We prove the result by showing that all the above variants of the 2D Block Tree, built on the matrix $B_k$ defined in Section \ref{sec:macroscheme}, contains $\Omega(N/\log^2 N)$ nodes. The proposition follows since by Lemma~\ref{lemma:gBk} it is $g(B_k) = O(\sqrt{N}\log\log N/\log N)$.
By Lemma~\ref{lemma:nok} until we reach the tree level in which the blocks are smaller than $k \times k$, all blocks are first occurrences, and therefore the corresponding tree nodes cannot be pruned. Hence, the 2D Block Tree nodes are $\Omega(n^2/k^2) = \Omega(N/\log^2 N)$. Note that the results hold even considering the variant described in~\cite[Theorem~4]{CarfagnaManzini2024} in which the number of nodes at the first level is $\Theta(\deltaCM(B_k))$, since, as shown in the proof of Proposition~\ref{prop:g_odeltaCM_square}, it is $\deltaCM(B_k)=\Omega(N/\log^2 N)$. 

For the attractor-based 2D Block Tree we note that in each level we mark the at least $\gammaCM$ nodes corresponding to blocks including an attractor position, and the result follows since $\gammaCM\geq\deltaCM$. Similarly, the result holds also for the variant described in~\cite[Theorem~5]{CarfagnaManzini2024}, since in that variant there are $\Theta(\gammaCM(B_k)) = \Omega(\deltaCM(B_k))$  nodes at the first level.\qed
\end{proof}}

\section{Effectiveness of Linearization Techniques}\label{sec:lin}

A classical heuristic for compressing 2D strings is to transform a matrix $\Mmn$ into a 1D string $S$ and use a one-dimensional compressor on~$S$. Having generalized 1D measures to 2D strings, it is natural to measure the effectiveness of linearization techniques by comparing, for a given measure $\mu$, the values $\mu(\Mmn)$ and $\mu(S)$. 
Clearly, for each matrix, there exists a linearization that makes the 2D string highly compressible: we can visit in order from left to right and from top to bottom all the occurrences of $a_1\in\Sigma$, followed by all the occurrences of $a_2\in\Sigma$, and so on, obtaining a string consisting in $|\Sigma|$ equal-letter runs.
However, this method requires an ad-hoc linearization for each matrix which may require substantial additional information to retrieve the original input. It is therefore customary in the literature to consider linearization techniques that can be inverted efficiently in terms of both time and space. 

\fullversion{The simplest linearization technique consists in mapping a matrix to the string obtained by concatenating its rows, as formally defined below.

\begin{definition}
The \emph{row-linearization} is the map $$\rowlin:\bigcup_{m,n>0}\Sigma^{m\times n}\mapsto \Sigma^*$$ such that $\rowlin(\Mmn)=\bigodot_{i=1}^{i=m}\MathM[i][1..n]=\MathM[1][1..n]\cdots \MathM[m][1..n]$. 
\end{definition}
}

The (lack of) effectiveness of \fullversion{$\rowlin$ } with respect to grammar compression has been already shown in~\cite[Theorem~2.2]{BermanKLPR02} with an example of a matrix $T_n$ of size $(2^{n+1}-1)\times (2^n+1)2^n$ such that $g(T_n) = O(n)$, while $g(\rowlin(T_n)) = \Omega(2^n)$. The following example shows a similar result for the measure~$\delta$.


\begin{example}\label{ex:delta_row_linearization}
    Let $\MathM_{n\times n}$ be obtained by appending to the identity matrix $I_{n-1}$ a row of $\zero$'s at the bottom, and then a column of $\one$'s at the right, as shown in Fig. \ref{fig:id01}. For each $k_1$, $k_2$, $P_\MathM(k_1, k_2)$ is at most $3(k_1+ k_2 )$. We can see this by considering three cases: the submatrices that do not intersect the last row or column, the submatrices intersecting the last row, and the submatrices intersecting the last column. 
    In each case, the distinct submatrices are associated to where the diagonal of $\one$'s intersects a submatrix (if it does so). This can happen in at most $k_1 + k_2$ different ways. As $3(k_1+k_2)/k_1k_2 \le 6$, we obtain $\delta(\Mnn)= O(1)$. On the other hand, for each $k \in [0\dd n-2]$ and $i \in [0\dd n-k-2]$, each factor $\zero^i\one\zero^k\one\zero^{n-k-i-2}$ appears in $\rowlin(\Mnn)$. There are $n-k-1$ of these factors for each $k$. Summing over all $k$, we obtain $P_\MathM(n)/n \geq (n-1)/2 = \Omega(n)$. Thus, $\delta(\rowlin(\Mnn))= \Omega(n)$.
\end{example}

\fullversion{
\begin{figure}[htb]
\centering
\begin{tikzpicture}[scale=0.75]
    [
        box/.style={rectangle,draw=black,thick, minimum size=1cm},
    ]

\foreach \x in {0,1,...,6}{
    \foreach \y in {0,-1,...,-6}{
         \ifthenelse{\x=-\y}{\node at (\x,\y){\one};}{\node at (\x,\y){\zero};}
}}
\foreach \x in {0,1,...,6}{
    \node at (\x,-7){\zero};
}
\foreach \y in {0,-1,...,-7}{
    \node at (7,\y){\one};
}

\draw[dashed] (0,0) to (7,0);
\draw[dashed] (7,0) to (0,-1);
\draw[dashed] (0,-1) to (7,-1);
\draw[dashed] (7,-1) to (0,-2);
\draw[dashed] (0,-2) to (7,-2);
\draw[dashed] (7,-2) to (0,-3);
\draw[dashed] (0,-3) to (7,-3);
\draw[dashed] (7,-3) to (0,-4);
\draw[dashed] (0,-4) to (7,-4);
\draw[dashed] (7,-4) to (0,-5);
\draw[dashed] (0,-5) to (7,-5);
\draw[dashed] (7,-5) to (0,-6);
\draw[dashed] (0,-6) to (7,-6);
\draw[dashed] (7,-6) to (0,-7);
\draw[dashed,->] (0,-7) to (6.9,-7);

\end{tikzpicture}
\caption{Example of the 2D string $\mathcal{M}_{8\times8}$ from Example~\ref{ex:delta_row_linearization}. The row-linearization is spelled by following the dashed arrow starting from the top-left corner.}
\label{fig:id01}
\end{figure}
}

Somewhat surprisingly, in some settings, the linearized matrix has a smaller measure. The following example shows a family of 2D strings $E_k$ for which $\gamma(\rowlin(E_k))$ is asymptotically smaller than $\gamma(E_k)$.

\begin{example}\label{ex:gamma_linearization}
Consider the matrix $E_k$ having size $N=k\times 2^k$ of Proposition~\ref{prop:g_odelta}. We note that the $i$-th row of $E_k$ is the periodic string $(\zero^{(2^{i-1})}\one^{(2^{i-1})})^{(2^{k-i})}$ and therefore $\rowlin(E_k)=\bigodot_{i=1}^{i=k} (\zero^{(2^{i-1})}\one^{(2^{i-1})})^{(2^{k-i})}$. 
We define the set $A=\bigcup_{i=1}^{i=k}\{(i-1)2^{k}+1,(i-1)2^{k}+1+2^{i-1},(i-1)2^{k}+2^i\}$, that is the set of positions of $\rowlin(E_k)$ where respectively the first $\zero$ and the first/last $\one$ of the leftmost occurrence of $\zero^{(2^{i-1})}\one^{(2^{i-1})}$ in row $i$ are mapped during the linearization. We claim that $A$ is an attractor for $\rowlin(E_k)$. If a substring $S$ of $\rowlin(E_k)$ spans more than one row of $E_k$, it includes a $\zero$ from the first column of $E_k$, and therefore it crosses a position of $A$. Otherwise $S$ is a substring of the $i$th row and since the rows of $E_k$ are periodic, the leftmost occurrence $S'$ of $S$ starts inside the first group of $\zero^{(2^{i-1})}\one^{(2^{i-1})}$ i.e. in $E_k[i][1..2^i]$. Suppose that $S'$ does not include any attractor position. Then, $S'$ has to be shorter than the maximum distance between two adjacent attractor positions in the same row i.e. it must be  $l=|S'|\leq 2^{i-1}-1$, and therefore $S'=\zero^a\one^{l-a}$ or $S'=\one^a\zero^{l-a}$ for some $0 \leq a < l$ because $S'$ cannot overlap two distinct groups of $\one$'s or $\zero$'s. If $a=0$ then $S'$ must include respectively the first $\one$ or the first $\zero$ in the run, otherwise it must include respectively the first or the last $\one$, therefore we conclude that $A$ is an attractor for $\rowlin(E_k)$.

Since $A$ has size $3k-1$, it is $\gamma(\rowlin(E_k))=O(k)$, on the other hand since each column of $E_k$ is a distinct non-overlapping $k\times1$ factor it is $\gamma(E_k)\geq 2^k$.
\end{example}

\fullversion{
\begin{figure}[htb]
\centering
\begin{tikzpicture}[scale=0.70]
    [
        box/.style={rectangle,draw=black,thick, minimum size=1cm},
    ]

\foreach \x in {0,1,...,15}{
    \ifthenelse{\x>7}{\node at (\x,-3){\one};}{\node at (\x,-3){\zero};}
}
\node at (0,-2){\zero};
\node at (1,-2){\zero};
\node at (2,-2){\zero};
\node at (3,-2){\zero};
\node at (4,-2){\one};
\node at (5,-2){\one};
\node at (6,-2){\one};
\node at (7,-2){\one};
\node at (8,-2){\zero};
\node at (9,-2){\zero};
\node at (10,-2){\zero};
\node at (11,-2){\zero};
\node at (12,-2){\one};
\node at (13,-2){\one};
\node at (14,-2){\one};
\node at (15,-2){\one};

\node at (0,-1){\zero};
\node at (1,-1){\zero};
\node at (2,-1){\one};
\node at (3,-1){\one};
\node at (4,-1){\zero};
\node at (5,-1){\zero};
\node at (6,-1){\one};
\node at (7,-1){\one};
\node at (8,-1){\zero};
\node at (9,-1){\zero};
\node at (10,-1){\one};
\node at (11,-1){\one};
\node at (12,-1){\zero};
\node at (13,-1){\zero};
\node at (14,-1){\one};
\node at (15,-1){\one};

\foreach \x in {0,1,...,15}{
    \ifthenelse{\isodd{\x}}{\node at (\x,0){\one};}{\node at (\x,0){\zero};}
}

\draw[dashed] (0,0) to (15,0);
\draw[dashed] (15,0) to (0,-1);
\draw[dashed] (0,-1) to (15,-1);
\draw[dashed] (15,-1) to (0,-2);
\draw[dashed] (0,-2) to (15,-2);
\draw[dashed] (15,-2) to (0,-3);
\draw[dashed,->] (0,-3) to (14.9,-3);

\end{tikzpicture}
\caption{Example of the 2D string $E_4$ from Example~\ref{ex:gamma_linearization}. The row-linearization is spelled by following the dashed arrow starting from the top-left corner.}
\label{fig:Ek}
\end{figure}
}

\fullversion{
Another well-known linearization technique uses a plane-filling curve, such as the Peano-Hilbert curve. Unlike the linearization discussed above, this is defined on square matrices whose dimensions are a power of 2. For such a definition, the notion of upper left $(UL)$, upper right $(UR)$, lower left $(LL)$, and lower right $(LR)$ quadrants is used~\cite{Lempel_ZIv_2D}. More formally, given the matrix $\MathM\in \Sigma^{2^i\times 2^i}$, $UL(\MathM)=\MathM[1\dd 2^{i-1}][1\dd 2^{i-1}]$, $UR(\MathM)=\MathM[1 \dd 2^{i-1}][2^{i-1}+1 \dd 2^i]$, $LL(\MathM)=\MathM[2^{i-1}+1 \dd 2^i][1 \dd 2^{i-1}]$, $LR(\MathM)=\MathM[2^{i-1}+1 \dd 2^i][2^{i-1}+1 \dd 2^i]$.

\begin{definition}
    The \emph{left, right, up and down scans}, denoted by $\LS$, $\RS$, $\US$, and $\DS$, respectively,  are the maps 
    $\bigcup_{i\geq0}\Sigma^{2^i\times 2^i}\rightarrow \Sigma^*$ defined recursively as follows:
    \begin{itemize}
        \item $\LS(\MathM) = \RS(\MathM) = \US(\MathM) = \DS(\MathM) = \MathM$, if $|\MathM|=1$;
        \item $\RS(\MathM) = \DS(UL(\MathM))\cdot \RS(UR(\MathM))\cdot \RS(LR(\MathM))\cdot \US(LL(\MathM))$, if $|\MathM|>1$;
        \item $\DS(\MathM) = \RS(UL(\MathM))\cdot \DS(LL(\MathM))\cdot \DS(LR(\MathM))\cdot \LS(UR(\MathM))$, if $|\MathM|>1$;
        \item $\US(\MathM) = \LS(LR(\MathM))\cdot \US(UR(\MathM))\cdot \US(UL(\MathM))\cdot \RS(LL(\MathM))$, if $|\MathM|>1$;
        \item $\LS(\MathM) = \US(LR(\MathM))\cdot \LS(LL(\MathM))\cdot \LS(UL(\MathM))\cdot \DS(UR(\MathM))$, if $|\MathM|>1$.
    \end{itemize}

    The \emph{Peano-Hilbert-linearization} is the map $$\phlin:\bigcup_{i\geq 0}\Sigma^{2^i\times 2^i}\rightarrow \Sigma^*$$ such that, for each matrix $\MathM$ of size $2^i\times 2^i$,
    \begin{itemize}
        \item $\phlin(\MathM)=\RS(\MathM)$ if $i$ is odd;
        \item $\phlin(\MathM)=\DS(\MathM)$ if $i$ is even.
    \end{itemize} 
\end{definition}
}

\fullversion{
From the definition of $\phlin$, it easily follows that each quadrant of a matrix $\MathM_{2^i\times2^i}$ is completely visited before moving to the following.
An example of Peano-Hilbert linearization of the identity matrix $I_{2^k}$, with $k=3$, is shown in Fig. \ref{fig:figph}.

The following two lemmas show that the right and down scans produce the same string when applied to the identity matrix (Lemma \ref{le:ds=rs}) and provide a characterization of the string $\phlin(I_{2^k})$ (Lemma \ref{le:phlin_id}).

\begin{lemma}
\label{le:ds=rs}
    Let us denote by $I_{2^k}$ the identity matrix of size $2^k\times 2^k$. Then, $\DS(I_{2^k}) = \RS(I_{2^k})$, for all $k\geq 0$. 
\end{lemma}
\begin{proof}
    We provide a proof by induction.
    For $k=0$, clearly $\DS(I_{2^0}) = \RS(I_{2^0}) = 1$.
    For the inductive step, assume that $\DS(I_{2^j}) = \RS(I_{2^j})$ for all $j\leq k$.
    For $t\geq0$, let $\zero_{2^t}$ denote the matrix of size $2^t\times2^t$ containing only $\zero$'s.
    Observe that $\DS(\zero_{2^t}) = \RS(\zero_{2^t}) = \LS(\zero_{2^t}) = \US(\zero_{2^t}) = \zero^{4^{t}}$, for all $t\geq 0$.
    Then,
    \begin{align*}
    \DS(I_{2^{k+1}})& = \RS(UL(I_{2^{k+1}}))\cdot \DS(LL(I_{2^{k+1}}))\cdot \DS(LR(I_{2^{k+1}}))\cdot \LS(UR(I_{2^{k+1}}))\\
    & = \RS(I_{2^{k}})\cdot \DS(\zero_{2^{k}})\cdot \DS(I_{2^{k}})\cdot \LS(\zero_{2^{k}})\\
    & = \DS(I_{2^{k}})\cdot \RS(\zero_{2^{k}})\cdot \RS(I_{2^{k}})\cdot \US(\zero_{2^{k}})\\
    & = \DS(UL(I_{2^{k+1}}))\cdot \RS(UR(I_{2^{k+1}}))\cdot \RS(LR(I_{2^{k+1}}))\cdot \US(LL(I_{2^{k+1}}))\\
    &= \RS(I_{2^{k+1}}),
    \end{align*}
    and the thesis follows.\qed
\end{proof}
}

\fullversion{
\begin{lemma}
\label{le:phlin_id}
For the identity matrix $I_{2^k}$ of size $2^k\times 2^k$, with $k\geq 1$ it holds that, 
\begin{equation*}
    \phlin(I_{2^{k}}) \;=\; 
    \phlin(I_{2^{k-1}})\texttt{0}^{4^{k-1}}
    \phlin(I_{2^{k-1}})\texttt{0}^{4^{k-1}}.
\end{equation*}
\end{lemma}

\begin{proof}
By Lemma \ref{le:ds=rs} it is $\phlin(I_{2^k}) = \RS(I_{2^k}) = \DS(I_{2^k})$ independently of the value of $k$, therefore for every $k\geq1$ it holds that
\begin{align*}
\phlin(I_{2^{k}})&= \RS(I_{2^k})\\
&=\DS(UL(I_{2^{k}}))\cdot \RS(UR(I_{2^{k}}))\cdot \RS(LR(I_{2^{k}}))\cdot \US(LL(I_{2^{k}}))\\
&=\DS(I_{2^{k-1}})\cdot \RS(0_{2^{k-1}})\cdot \RS(I_{2^{k-1}})\cdot \US(0_{2^{k-1}})\\ 
&=\phlin(I_{2^{k-1}})\cdot \texttt{0}^{4^{k-1}}\cdot \phlin(I_{2^{k-1}})\cdot \texttt{0}^{4^{k-1}}
\end{align*}\qed 
\end{proof}
}

Lempel and Ziv~\cite[Lemma 1]{Lempel_ZIv_2D} showed that $\phlin$ is effective for compressing 2D strings using a finite-state encoder. 
In the next proposition, we show that there exist matrices with a 2D bidirectional macro scheme that is smaller by a logarithmic factor compared to the optimal scheme of their $\phlin$ linearized form.



\begin{proposition}
    It is $b(I_{2^k})=O(1)$ and $b(\phlin(I_{2^k})) = \Omega(k)$.
\end{proposition}

\begin{proof}
    We have already observed in Example~\ref{ex:b_O1} that  $b(I_n)=O(1)$ for any $n>0$.
    To prove the second bound, for $\ell\geq 1$ define $t_\ell=\sum_{i=0}^{\ell-1} 4^i$. We preliminary prove by induction on~$k$ that: $a$) $\phlin(I_{2^k})$ starts with $\one$, $b$) $\phlin(I_{2^k})$ ends with $\one \zero^{t_k}$, $c$) $\phlin(I_{2^k})$ contains the substrings $\one\zero^{t_\ell}\one$ for $\ell=1,\dd,k$. 
    
    For $k=1$ $\phlin(I_{2})=\texttt{1010}$ satisfies all conditions. For the inductive step $a$) is trivial. By Lemma \ref{le:phlin_id}, $\phlin(I_{2^{k+1}})$ ends with $\phlin(I_{2^k})\zero^{4^k}$, then $b$) is immediate. To prove $c$) observe that, by Lemma \ref{le:phlin_id}, $\phlin(I_{2^{k+1}})$ contains $\phlin(I_{2^k})$. Then, by induction it contains all substrings $\one\zero^{t_\ell}\one$ for $\ell=1,\dd,k$. To prove that it also contains $\one\zero^{t_{k+1}}\one$ observe that by $a$) $\phlin(I_{2^{k+1}})$ starts with $\phlin(I_{2^k})\zero^{4^k} \one$ and by $b$) this string ends with $\one \zero^{t_k}\zero^{4^k} \one = \one \zero^{t_{k+1}}\one$, and therefore is a substring of $\phlin(I_{2^{k+1}})$. 

    Having established that $\phlin(I_{2^k})$ contains the {\em distinct} substrings $\one\zero^{t_\ell}\one$ for $\ell=1,\dd,k$,  since a single string position can be contained in at most two such substrings, we conclude that $b(\phlin(I_{2^k})) \geq \gamma(\phlin(I_{2^k})) = \Omega(k)$. \qed
\end{proof}

\fullversion{
\begin{figure}[hbt]
\centering
\begin{tikzpicture}
    [
        box/.style={rectangle,draw=black,thick, minimum size=1cm},
    ]

\foreach \x in {0,1,...,7}{
    \foreach \y in {0,-1,...,-7}{
         \ifthenelse{\x=-\y}{\node at (\x,\y){\one};}{\node at (\x,\y){\zero};}
}}
\draw[dashed] (0,0) to (1,0);
\draw[dashed] (1,0) to (1,-1);
\draw[dashed] (1,-1) to (0,-1);
\draw[dashed] (0,-1) to (0,-3);
\draw[dashed] (0,-3) to (1,-3);
\draw[dashed] (1,-3) to (1,-2);
\draw[dashed] (1,-2) to (2,-2);
\draw[dashed] (2,-2) to (2,-3);
\draw[dashed] (2,-3) to (3,-3);
\draw[dashed] (3,-3) to (3,-1);
\draw[dashed] (3,-1) to (2,-1);
\draw[dashed] (2,-1) to (2,0);
\draw[dashed] (2,0) to (4,0);
\draw[dashed] (4,0) to (4,-1);
\draw[dashed] (4,-1) to (5,-1);
\draw[dashed] (5,-1) to (5,0);
\draw[dashed] (5,0) to (7,0);
\draw[dashed] (7,0) to (7,-1);
\draw[dashed] (7,-1) to (6,-1);
\draw[dashed] (6,-1) to (6,-2);
\draw[dashed] (6,-2) to (7,-2);
\draw[dashed] (7,-2) to (7,-3);
\draw[dashed] (7,-3) to (5,-3);
\draw[dashed] (5,-3) to (5,-2);
\draw[dashed] (5,-2) to (4,-2);
\draw[dashed] (4,-2) to (4,-5);
\draw[dashed] (4,-5) to (5,-5);
\draw[dashed] (5,-5) to (5,-4);
\draw[dashed] (5,-4) to (7,-4);
\draw[dashed] (7,-4) to (7,-5);
\draw[dashed] (7,-5) to (6,-5);
\draw[dashed] (6,-5) to (6,-6);
\draw[dashed] (6,-6) to (7,-6);
\draw[dashed] (7,-6) to (7,-7);
\draw[dashed] (7,-7) to (5,-7);
\draw[dashed] (5,-7) to (5,-6);
\draw[dashed] (5,-6) to (4,-6);
\draw[dashed] (4,-6) to (4,-7);
\draw[dashed] (4,-7) to (2,-7);
\draw[dashed] (2,-7) to (2,-6);
\draw[dashed] (2,-6) to (3,-6);
\draw[dashed] (3,-6) to (3,-4);
\draw[dashed] (3,-4) to (2,-4);
\draw[dashed] (2,-4) to (2,-5);
\draw[dashed] (2,-5) to (1,-5);
\draw[dashed] (1,-5) to (1,-4);
\draw[dashed] (1,-4) to (0,-4);
\draw[dashed] (0,-4) to (0,-6);
\draw[dashed] (0,-6) to (1,-6);
\draw[dashed] (1,-6) to (1,-7);
\draw[dashed, ->] (1,-7) to (0.1,-7);
\end{tikzpicture}
\caption{Example of the 2D string $I_8$. The Peano-Hilbert linearization is spelled by following the dashed arrow starting from the top-left corner. The string $\phlin(I_8)$ is obtained as $\RS(I_8)=\DS(UL(I_8))\cdot \RS(UR(I_8))\cdot \RS(LR(I_8))\cdot \US(LL(I_8))=\phlin(I_4)\cdot 0^{16}\cdot \phlin(I_4)\cdot 0^{16}$.}
\label{fig:figph}
\end{figure}
}

\fullversion{

\section{Generalization to Multidimensional Strings}
In the previous sections, we focused on extending repetitiveness measures of strings to the two-dimensional context.
Here, we outline how the definitions of the measures $\delta$, $\gamma$, $g_{rl}$, $g$, and $b$, and some of their properties, can be naturally generalized to $\Dim$-dimensional strings, for every $\Dim>0$. Note that the measures $\gammaCM$, $\deltaCM$ have been generalized to 3D strings in~\cite{CarfagnaManzini2024}, where it is shown that in 3D it is still $\deltaCM \leq \gammaCM$ and that the gap between them can be larger than in 2D. 

Given a tuple of $\Dim$ integers $\mathbf{n} = (n_1,  \ldots, n_\Dim)$, with $n_i> 0$ for all $i\in[1\dd \Dim]$, a \emph{$\Dim$D string} $\MathM_{\mathbf{n}}$ is a multidimensional array of elements from $\Sigma$ of size $N = n_1 \times \stackrel{\Dim}{\cdots} \times n_\Dim$, where $n_i$ is the size over the $i$th dimension.
Every element of $\MathMM$ can be accessed by $\MathMM[\mathbf{j}]\in\Sigma$, where $\mathbf{j}$ is a $\Dim$-tuple $\mathbf{j} = (j_1, \ldots, j_\Dim)$, with $1 \leq j_i \leq n_i$ for all $i\in[1\dd \Dim]$, or alternatively using the notation $\MathMM[j_1]..[j_\Dim]$.

Given two positions $\mathbf{i},\mathbf{j}$ of $\MathMM$, we denote by $\MathMM[\mathbf{i}\dd\mathbf{j}]$ the $\Dim$D substring starting at position $\mathbf{i}$ and ending at position $\mathbf{j}$, that is $\MathM[i_1\dd j_1][i_2 \dd j_2]\cdots[i_\Dim\dd j_\Dim]$.
Given two $\Dim$D strings $\MathMM$ and $\MathM'_{\mathbf{m}}$, for some $\mathbf{n} = (n_1, \ldots, n_\Dim)$ and $\mathbf{m} = (m_1, \ldots, m_\Dim)$, the concatenation over the $k$th dimension of $\MathMM$ and $\MathM'_{\mathbf{m}}$, denoted by $\MathMM\ohrz_{(k)} \MathM'_{\mathbf{m}}$, is a partial operation that can be performed
only if $n_j = m_j$ for all $j\neq k$. This operation produces a $\Dim$D string with the same size of $\MathMM$ and $\MathM'_{\mathbf{m}}$ along every direction except along the $k$th one where the resulting size becomes $n_k+m_k$.

Formally, the resulting matrix is defined as follows: given a $\Dim$-tuple $\mathbf{i} = (i_1, \ldots, i_\Dim)$ such that $i_j\in[1\dd n_j]$ when $j\neq k$ and $i_k\in[1\dd n_k+m_k]$, it is:
\begin{itemize}
    \item ($\MathMM\ohrz_{(k)} \MathM'_{\mathbf{m}})[\mathbf{i}]=\MathMM[\mathbf{i}]$ if $i_k \leq n_k$;
    \item ($\MathMM\ohrz_{(k)} \MathM'_{\mathbf{m}})[\mathbf{i}]=\MathM'_\mathbf{m}[\mathbf{i'}]$ otherwise
\end{itemize}

where $\mathbf{i'}$ is a $\Dim$-tuple with ${i_j}' = i_j$ for all $j\neq k$ and $i'_k = i_k-n_k$.}

\fullversion{

The $\Dim$D substring complexity $P_\MathM$ counts for each $\Dim$-tuple of positive integers $(k_1,\ldots, k_\Dim)$ the number of distinct $(k_1 \times \cdots \times k_\Dim)$-factors in $\MathM_\mathbf{n}$. 

In the following, we provide the above-mentioned definitions of the measures $\delta$, $\gamma$, $g$, $g_{rl}$, and $b$ for the $\Dim$-dimensional case. Observe that as $d=2$, these match with the definitions given in the previous sections. 

\begin{definition} \label{def:Mddelta}Let $\MathM_\mathbf{n}$ be a $\Dim$D string, for some $\Dim>0$ and $\mathbf{n} = (n_1, \ldots, n_\Dim)$, and let $P_\MathM$ be the $\Dim$D substring complexity of $\MathM_\mathbf{n}$.
Then, $$\delta(\MathM_\mathbf{n}) = \max\left\{\frac{P_\MathM(k_1, \ldots, k_\Dim)}{k_1\cdots k_\Dim}, 1\leq k_i\leq n_i \text{ for all } 1\leq i\leq \Dim\right\}.$$
\end{definition} 

\begin{definition}\label{def:Mdstrinattractor}An  \emph{attractor} for a $\Dim$D string $\MathMM$, for some $\Dim>0$ and $\nn = (n_1, \ldots, n_\Dim)$, is a set $\Gamma \subseteq [1\dd n_1] \times [1\dd n_2] \times \cdots \times [1 \dd n_\Dim]$ with the property that any $\Dim$D substring $\MathM[i_1\dd j_1][i_2 \dd j_2]\cdots[i_\Dim\dd j_\Dim]$ of $\MathMM$ has an occurrence $\MathM[i'_1\dd j'_1][i'_2 \dd j'_2]\cdots[i'_\Dim\dd j'_\Dim]$ such 
that $\exists \mathbf{x}=(x_1, \ldots, x_\Dim) \in \Gamma$ with $i_i' \le x_i \le j'_i$ for all $i\in[1\dd \Dim]$. The size of the smallest attractor for $\MathMM$ is denoted by $\gamma(\MathMM)$. 
\end{definition}

\begin{definition}
Let $\MathMM$ be a $\Dim$D string, for some $\Dim>0$ and $\mathbf{n} = (n_1, \ldots, n_\Dim)$. A \emph{$\Dim$-dimensional Straight-Line Program} ($\Dim$D SLP) for $\MathMM$ is a context-free grammar $(V, \Sigma, R, S)$ that uniquely generates $\MathMM$ and where the definition of the right-hand side of a variable can have the form $$A \rightarrow a,\, A \rightarrow B\ohrz_{(i)} C, \text{ with } i\in[1\dd\Dim],$$ where $a \in \Sigma$, $B, C \in V$. We call these definitions \emph{terminal rules} and \emph{$i$-rules} respectively. The expansion of a variable is defined as $$\gexp(A) = a,\, \gexp(A) = \gexp(B)\ohrz_{(i)} \gexp(C), \text{ with } i\in[1\dd\Dim],$$ respectively.

The size $|G|$ of a $\Dim$D SLP $G$ is the sum of the sizes of all the rules of $G$, where we assume that the terminal rules have size $1$ and the $i$-rules have size $2$.
The measure $g(\MathMM)$ is defined as the size of the smallest $\Dim$D SLP generating $\MathMM$.

\end{definition}

\begin{definition}
A \emph{$\Dim$-dimensional Run-Length Straight-Line program} ($\Dim$D RLSLP) is a $\Dim$D SLP that in addition allows special rules, which are assumed to be of size 2, of the form
$$A \rightarrow  \ohrz^k_{(i)} B,  \text{ with } i\in[1\dd \Dim], \text{ and } k>1$$ 
with their expansions defined as 
\begin{align*} 
    \gexp(A) =  \underbrace{\gexp(B)\ohrz_{(i)} \gexp(B) \ohrz_{(i)} \cdots \ohrz_{(i)} \gexp(B)}_{k \text{ times}}
\end{align*}
The measure $g_{rl}(\MathMM)$ is defined as the sum of the size of the rules of a smallest $\Dim$D RLSLP generating $\MathMM$.
\end{definition}

\begin{definition}\label{def:macroD}
Let $\MathMM$ be a $\Dim$D string, for some $\Dim>0$ and $\mathbf{n} = (n_1, \ldots, n_\Dim)$.
A \emph{$\Dim$D macro scheme} for $\MathMM$ is any factorization of $\MathMM$ into a set of disjoint phrases such that any phrase is either an element from $\Sigma$ called an \emph{explicit symbol/phrase}, or is a copied phrase with source in $\MathMM$ starting at a different position. For a $\Dim$D macro scheme to be \emph{valid} or \emph{decodable}, the function $$\map: ([1 \dd n_1] \times \cdots \times [1\dd n_\Dim])  \cup \{\bot\} \rightarrow ([1 \dd n_1] \times [1\dd n_{\Dim}]) \cup \{\bot\}$$ induced by the factorization must verify that: 
\begin{enumerate}
    \item[i)] $\map(\bot) = \bot$, and if $\MathMM[\mathbf{i}]$ is an explicit symbol, then $\map(\mathbf{i}) = \bot$;
    \item[ii)] for each copied phrase $\MathMM[\mathbf{i}\dd \mathbf{j}]$, it must hold that $\map(\mathbf{i}+\mathbf{t}) = \map(\mathbf{i}) + \mathbf{t}$ for $\mathbf{t} \in [0\dd j_1-i_1] \times \cdots \times [0\dd j_\Dim - i_\Dim]$, where $\map(\mathbf{i})$ is the top-left corner of the source for $\MathMM[\mathbf{i}\dd \mathbf{j}]$;
    \item[iii)] for each $\mathbf{i} \in [1\dd n_1] \times \cdots \times [1 \dd n_\Dim]$ there exists $k > 0$ such that $\map^k(\mathbf{i}) = \bot$. 
\end{enumerate}
We define $b(\MathMM)$ as the smallest number of phrases in a valid $\Dim$D macro scheme for $\MathMM$.
\end{definition}

It is not hard to see that Propositions~\ref{prop:delta<=gamma}, \ref{prop:grl<=g}, and~\ref{prop:b<=grl}, establishing that $\delta\leq\gamma$, and  $b \leq g_{rl} \leq g$, can be generalized to the $\Dim$-dimensional case, with analogous proofs, for all $\Dim>2$.
We conclude this section by showing a class of multidimensional strings for which the gap between the measures $\delta$ and $g$ grows with the number of dimensions $\Dim$.

Consider again a binary de Bruijn sequence $D_k$ of size $n = 2^k +k -1$.
We define the $\Dim$D string $B_{\Dim,k}$ of size $N=\overbrace{n\times \cdots \times n}^\Dim$ over the alphabet $\Delta_\Dim=\{\langle b_1,\ldots,b_\Dim \rangle \mid b_i\in \{\zero,\one \}, i\in [1\dd \Dim]\}$ of size $2^\Dim$ by the following relation: given a position $\mathbf{i} = (i_1, \ldots, i_\Dim)\in[1\dd n]^\Dim$ it is $$B_{\Dim,k}[\mathbf{i}] = \langle D_k[i_1], \ldots, D_k[i_\Dim]\rangle.$$ 
Observe that the definition of $B_{\Dim,k}$ is a generalization of the 2D string $B_k$ defined in Section~\ref{sec:macroscheme} since $B_{2,k} = B_k$.

\begin{lemma}\label{le:gBdk}
    For every $\Dim,k>0$ it is $g(B_{\Dim,k}) = O(2^\Dim n\log\log n/\log n)$.
\end{lemma}

\begin{proof}
For $\Dim=1$, Gagie et al. showed in~\cite{latin/GagieNP18} that the statement is true for all $k$, that is, there exists a one-dimensional SLP $G_{1,k}$ of size $O(n\log\log n/\log n)$ generating the de Bruijn sequence $D_k$. In the following, we exhibit a $\Dim$-dimensional SLP $G_{\Dim,k}$ of the claimed size which expands to $B_{\Dim,k}$ for all $\Dim$ and $k$. 
We note that for every $i$, the substring $B_{\Dim,k}[1..n]..[1..n][i]$ can be of only two types depending on the value of $D_k[i]$, in particular, it is equal (comparing them element by element) to $B_{\Dim-1,k}$ where we replaced every element $B_{\Dim-1,k}[i_1]..[i_{d-1}]=\langle D_k[i_1], \ldots, D_k[i_{\Dim-1}]\rangle$ with respectively $\langle D_k[i_1], \ldots, D_k[i_{\Dim-1}],\zero \rangle$ if $D_k[i]=\zero$, or with $\langle D_k[i_1], \ldots, D_k[i_{\Dim-1}],\one \rangle$ otherwise. In the following we call these two $(\Dim-1)$D strings respectively $B^\zero_{\Dim-1,k}$ and $B^\one_{\Dim-1,k}$.
We obtain a grammar $G_{\Dim,k}$ for $B_{\Dim,k}$ inductively as follows. Given a grammar $G_{\Dim-1,k}$ for $B_{\Dim-1,k}$ we obtain the grammars $G^{\zero}_{\Dim-1,k}$ and $G^{\one}_{\Dim-1,k}$ corresponding to $B^\zero_{\Dim-1,k}$ and $B^\one_{\Dim-1,k}$ by replacing in $G_{\Dim-1,k}$ each terminal symbol $\langle b_1,\ldots,b_{\Dim-1}\rangle$ with $\langle b_1,\ldots,b_{\Dim-1},\zero\rangle$ or $\langle b_1,\ldots,b_{\Dim-1},\one\rangle$ respectively; in both cases without changing the size of the grammars. To obtain the final grammar we transform all the horizontal rules of the one-dimensional grammar $G_{1,k}$ in $\ohrz_{(d)}$ rules, and we replace the terminals $\zero$ and $\one$ respectively with the starting symbols of the grammars $G^{\zero}_{\Dim-1,k}$ and $G^{\one}_{\Dim-1,k}$ without changing the size of $G_{1,k}$. 
As a consequence the resulting grammar has size $|G_{\Dim,k}|= O(n\log\log n/\log n) + 2|G_{\Dim-1,k}| = O(2^\Dim n\log\log n/\log n)$. \qed

\end{proof}

\medskip

The proof of Lemma~\ref{lemma:nok} can be easily generalized to derive the following.
\begin{lemma}\label{lemma:NOK}
    Every substring of size $\underbrace{k\times\cdots\times k}_\Dim$ appears in $B_{\Dim,k}$ at most once.
\end{lemma}
\begin{proof}
We observe that by definition of $B_{\Dim,k}$, moving in $B_{\Dim,k}$ along every fixed dimension $j$, we 
read a de Bruijn sequence over a binary alphabet. More formally, 
for all $j$, the substring $B_{\Dim,k}[i_1]..[i_{j-1}][1..n][i_{j+1}]..[i_{\Dim}]$ spells a de Bruijn sequence of order $k$ over the binary alphabet having symbols \sloppy $\langle D_k[i_1], .., D_k[i_{j-1}],\zero,D_k[i_{j+1}],..,D_k[i_\Dim]\rangle$ and $\langle D_k[i_1], .., D_k[i_{j-1}],\one,D_k[i_{j+1}],..,D_k[i_\Dim]\rangle$. As a consequence, there cannot be two distinct occurrences of the same $\underbrace{k\times {\cdots} \times k}_\Dim$ substring.  \qed
\end{proof}

\begin{proposition} \label{prop:g_odeltaMultidim}
For every $\Dim>0$, there exists an infinite family of $\Dim$D strings with $\delta = \Omega\left(\frac{g N^{\frac{\Dim-1}{\Dim}}}{\log\log N}\left(\frac{\Dim}{2\log N}\right)^{\Dim-1}\right)$, where $N$ is the size of the input string. 
\end{proposition}
\begin{proof}    
    Consider the $\Dim$D string $B_{\Dim,k}$, for some $\Dim,k\geq1$.
    By Lemma~\ref{le:gBdk} it is $g(B_{\Dim,k}) = O(2^\Dim n\log\log n/\log n) = O(\Dim 2^\Dim N^{\frac{1}{\Dim}}\log\log N/\log N)$. 
    By Lemma~\ref{lemma:NOK} $B_{\Dim,k}$ contains $\Theta(n^\Dim) = \Theta(N)$ distinct $\underbrace{k\times {\cdots} \times k}_\Dim$ factors, hence  $\delta = \Omega\left(\frac{N\Dim^\Dim}{\log^\Dim N}\right)$.\qed
\end{proof}

 }



\fullversion{
\subsection{Direct access to $d$D RLSLPs}

The heavy-path approach for direct access on 2D RLSLPs can be generalized to $d$D RLSLPs. We need to take into account that in $d$D RLSLPs, there are $d$ predecessor queries that need to be made to change the heavy-path. 
In terms of space, for each variable we need to store its dimension array (the size of its expansion), the coordinate of its heavy symbol (the heavy occurrence), and $\Dim$ pair of size sequences values. 

After making the above considerations, the following result is straightforward.
 
\begin{theorem}\label{thm:direct_access_grl_D}Let $\MathMM$ be a $N$-size $d$D string and let $G_{rl}$ be a $d$D RLSLP generating $\MathMM$, for some $\mathbf{n} = (n_1,\ldots, n_\Dim)$. There exists a data structure that uses 
$O(|G_{rl}|\log N)$ bits of space and supports direct access queries to any cell \sloppy $\MathMM[\mathbf{i}]$ in $O(\log  N)$
time, by assuming that $\Dim$ processes can be executed in parallel.
\end{theorem}
}

\section{Conclusions and Future Works}

In this paper, we have shown how to generalize the 
repetitiveness measures previously used in the one-dimensional context to generic two-dimensional strings. In particular, we have introduced extensions to the 2D case of the measures $\delta$ and $\gamma$ based on distinct factors of arbitrary rectangular shape, as well as the extensions of the measures $g$, $g_{rl}$, and $b$, which are based on copy-paste mechanisms.
We have studied the mutual relationships between these measures and we have shown that $\delta\leq \gamma$ and $b\leq g_{rl}\leq g$.  We have proven that, unlike in the 1D context where $\delta\leq \gamma \leq b\leq g_{rl}\leq g$, the two classes of measures become incomparable when 2D strings are considered. Indeed, we have shown that, depending on the 2D input, the measures $g$, $g_{rl}$, and $b$ can be asymptotically smaller than $\delta$ and $\gamma$.

The results presented in the paper highlight that in the 2D case, the measures $\delta$ and $\gamma$ (as well as their square-based versions introduced in \cite{CarfagnaManzini2023,CarfagnaManzini2024}) are not completely satisfactory for capturing the regularities of a generic two-dimensional string, which are instead effectively detected by $g$, $g_{rl}$, and $b$ measures. 

We have also analyzed the recently introduced 2D Block-Tree data structure~\cite{BrisaboaGGN24} which is able to compress a 2D string and provide efficient access to its individual symbols. Our results show that for some 2D strings the 2D Block-Tree fails to achieve a compression close to $g$ (and therefor $g_{rl}$ and $b$). 
We have also studied the use of linearization strategies as preprocessing to compress two-dimensional input, and shown that they are not always effective even when considering approaches based on the Peano-Hilbert space-filling curve. 

Our results indicate that the problem of finding a time-efficient 2D compression scheme that approaches the theoretically well-grounded measures $g$, $g_{rl}$, or $b$ is still open. A possible avenue to tackle this problem could be to explore possible approximation strategies for $b$ and $g$, as well as 2D versions of greedy grammar construction algorithms like the ones described in~\cite{Bannai_Hirayama_Hucke_Inenaga_Jez_Lohrey_Reh_2021,NOP20}. 
Our analysis suggests that another strategy worth pursuing is to modify the 2D-block tree so that it is not limited to searching for identical square substrings; handling rectangular shapes as well appears to be essential for maximizing compression.

\begin{credits}
\subsubsection{\ackname}
LC and GM are partially funded by the PNRR ECS00000017 Tuscany Health Ecosystem, Spoke 6, CUP I53C22000780001, funded by the NextGeneration EU programme,  by the spoke ``FutureHPC \& BigData'' of the ICSC --- Centro Nazionale di Ricerca in High-Performance Computing, Big Data and Quantum Computing, funded by the NextGeneration EU programme.

GR and MS are partially funded by the MUR PRIN Project \vir{PINC, Pangenome INformatiCs: from Theory to Applications} (Grant No.\ 2022YRB97K).

MS is partially supported by Project ``ACoMPA'' (CUP B73C24001050001) funded by the NextGeneration EU programme PNRR ECS00000017 Tuscany Health Ecosystem (Spoke 6).

LC, GM, MS, and GR are partially funded by the INdAM-GNCS Project CUP E53C23001670001.

CU is partially funded \sloppy by ANID-Subdirección de Capital Humano/Doctorado Nacional/2021-21210580, ANID, Chile, partially funded by Basal Funds FB0001, ANID, Chile, and partially funded by Fondecyt Grant 1-230755.

\subsubsection{\discintname}
The authors have no competing interests to declare that are relevant to the content of this article.
    
\end{credits}

\bibliographystyle{plain}
\bibliography{bibliography.bib}

\end{document}